\documentclass{article}

\usepackage{latexsym}
\usepackage{amsfonts}
\usepackage{fullpage}
\usepackage[english]{babel}

\usepackage{amsthm}
\usepackage{color}

\input amssym.def
\input amssym.tex

\def\la{\mathrel{\mathchoice {\vcenter{\offinterlineskip\halign{\hfil
$\displaystyle##$\hfil\cr<\cr\noalign{\vskip1.5pt}\sim\cr}}}
{\vcenter{\offinterlineskip\halign{\hfil$\textstyle##$\hfil\cr<\cr
\noalign{\vskip1.0pt}\sim\cr}}}
{\vcenter{\offinterlineskip\halign{\hfil$\scriptstyle##$\hfil\cr<\cr
\noalign{\vskip0.5pt}\sim\cr}}}
{\vcenter{\offinterlineskip\halign{\hfil$\scriptscriptstyle##$\hfil
\cr<\cr\noalign{\vskip0.5pt}\sim\cr}}}}}

\def\utw{\smash{\rlap{\lower5pt\hbox{$\sim$}}}}
\def\udtw{\smash{\rlap{\lower6pt\hbox{$\approx$}}}}

\def\bbbone{{\mathchoice {\rm 1\mskip-4mu l} {\rm 1\mskip-4mu l}
{\rm 1\mskip-4.5mu l} {\rm 1\mskip-5mu l}}}
\def\bbbc{{\mathchoice {\setbox0=\hbox{$\displaystyle\rm C$}\hbox{\hbox
to0pt{\kern0.4\wd0\vrule height0.9\ht0\hss}\box0}}
{\setbox0=\hbox{$\textstyle\rm C$}\hbox{\hbox
to0pt{\kern0.4\wd0\vrule height0.9\ht0\hss}\box0}}
{\setbox0=\hbox{$\scriptstyle\rm C$}\hbox{\hbox
to0pt{\kern0.4\wd0\vrule height0.9\ht0\hss}\box0}}
{\setbox0=\hbox{$\scriptscriptstyle\rm C$}\hbox{\hbox
to0pt{\kern0.4\wd0\vrule height0.9\ht0\hss}\box0}}}}
\def\bbbe{{\mathchoice {\setbox0=\hbox{\smalletextfont e}\hbox{\raise
0.1\ht0\hbox to0pt{\kern0.4\wd0\vrule width0.3pt
height0.7\ht0\hss}\box0}} {\setbox0=\hbox{\smalletextfont
e}\hbox{\raise 0.1\ht0\hbox to0pt{\kern0.4\wd0\vrule width0.3pt
height0.7\ht0\hss}\box0}} {\setbox0=\hbox{\smallescriptfont
e}\hbox{\raise 0.1\ht0\hbox to0pt{\kern0.5\wd0\vrule width0.2pt
height0.7\ht0\hss}\box0}} {\setbox0=\hbox{\smallescriptscriptfont
e}\hbox{\raise 0.1\ht0\hbox to0pt{\kern0.4\wd0\vrule width0.2pt
height0.7\ht0\hss}\box0}}}}
\def\bbbq{{\mathchoice {\setbox0=\hbox{$\displaystyle\rm Q$}\hbox{\raise
0.15\ht0\hbox to0pt{\kern0.4\wd0\vrule height0.8\ht0\hss}\box0}}
{\setbox0=\hbox{$\textstyle\rm Q$}\hbox{\raise 0.15\ht0\hbox
to0pt{\kern0.4\wd0\vrule height0.8\ht0\hss}\box0}}
{\setbox0=\hbox{$\scriptstyle\rm Q$}\hbox{\raise 0.15\ht0\hbox
to0pt{\kern0.4\wd0\vrule height0.7\ht0\hss}\box0}}
{\setbox0=\hbox{$\scriptscriptstyle\rm Q$}\hbox{\raise
0.15\ht0\hbox to0pt{\kern0.4\wd0\vrule height0.7\ht0\hss}\box0}}}}
\def\bbbt{{\mathchoice {\setbox0=\hbox{$\displaystyle\rm
T$}\hbox{\hbox to0pt{\kern0.3\wd0\vrule height0.9\ht0\hss}\box0}}
{\setbox0=\hbox{$\textstyle\rm T$}\hbox{\hbox
to0pt{\kern0.3\wd0\vrule height0.9\ht0\hss}\box0}}
{\setbox0=\hbox{$\scriptstyle\rm T$}\hbox{\hbox
to0pt{\kern0.3\wd0\vrule height0.9\ht0\hss}\box0}}
{\setbox0=\hbox{$\scriptscriptstyle\rm T$}\hbox{\hbox
to0pt{\kern0.3\wd0\vrule height0.9\ht0\hss}\box0}}}}
\def\bbbs{{\mathchoice
{\setbox0=\hbox{$\displaystyle     \rm S$}\hbox{\raise0.5\ht0\hbox
to0pt{\kern0.35\wd0\vrule height0.45\ht0\hss}\hbox
to0pt{\kern0.55\wd0\vrule height0.5\ht0\hss}\box0}}
{\setbox0=\hbox{$\textstyle        \rm S$}\hbox{\raise0.5\ht0\hbox
to0pt{\kern0.35\wd0\vrule height0.45\ht0\hss}\hbox
to0pt{\kern0.55\wd0\vrule height0.5\ht0\hss}\box0}}
{\setbox0=\hbox{$\scriptstyle      \rm S$}\hbox{\raise0.5\ht0\hbox
to0pt{\kern0.35\wd0\vrule height0.45\ht0\hss}\raise0.05\ht0\hbox
to0pt{\kern0.5\wd0\vrule height0.45\ht0\hss}\box0}}
{\setbox0=\hbox{$\scriptscriptstyle\rm S$}\hbox{\raise0.5\ht0\hbox
to0pt{\kern0.4\wd0\vrule height0.45\ht0\hss}\raise0.05\ht0\hbox
to0pt{\kern0.55\wd0\vrule height0.45\ht0\hss}\box0}}}}

\def\bbbz{{\mathchoice {\hbox{$\sf\textstyle Z\kern-0.4em Z$}}
{\hbox{$\sf\textstyle Z\kern-0.4em Z$}} {\hbox{$\sf\scriptstyle
Z\kern-0.3em Z$}} {\hbox{$\sf\scriptscriptstyle Z\kern-0.2em
Z$}}}}

\def\diameter{{\ifmmode\oslash\else$\oslash$\fi}}

\newtheorem{teo}{Theorem}[section]
\newtheorem{prop}{Proposition}[section]
\newtheorem{Lemma}{Lemma}[section]

\newtheorem{Corollary}{Corollary}[section]

\theoremstyle{definition}
\newtheorem{rmk}{Remark}[section]

\def\proof{\noindent \textbf{Proof. }}

\newcommand {\atensorone}[2]{\bigwedge_{#1=1}^{#2}}

\newcommand {\dd}[2]{\frac{\partial{#1}}{\partial{#2}}}

\newcommand{\norme}[1]{\left\|#1 \right\|}
\newcommand{\module}[1]{|#1|}

\newcommand{\normebk}[1]{\|#1\|_{\B(\cK)}}
\newcommand{\normeH}[1]{\|#1\|_{\cH}}
\newcommand{\normeh}[1]{\|#1\|_{\ch}}

\def\Im{{\rm Im}}
\def\Re{{\rm Re}}

\def\rr{\mathbb{R}}
\def\rrt{\mathbb{R}^3}
\def\rrtn{\mathbb{R}^{3N}}

\def\C{\mathbb{C}}
\def\N{\mathbb{N}}

\def\H{\mathcal{H}}
\def\B{\mathcal{B}}
\def\BK{\mathcal{B}(\mathcal{K})}

\def\cH{{\cal H}}

\def\cK{{\cal K}}
\def\K{{\cal K}}
\def\ch{{\frak h}}
\def\D{{\cal D}}

\def\one{\bbbone}
\def\onea{\one_a}
\def\oneac{\one_{a^C}}

\def\cN{{\rm N}}

\def\d{{\rm d}}

\def\la{\lambda}
\def\La{\Lambda}
\def\eps{\varepsilon}
\def\G{\Gamma}

\def\12{1/2}
\def\32{3/2}

\def\om{\omega}

\def\Om{\Omega}

\def\Si{\Sigma}
\def\si{\sigma}

\def\ic{{{\rm i}}}

\def\ot{\otimes}
\def\otc{\hat{\otimes}}
\def\nparts{\bigotimes_{\rm s}^n \frak{h}}
\def\npartsuno{\bigotimes_{\rm s}^n \frak{h}_1}

\def\l2{{L}^2}
\def\antisl2{\bigwedge^N \rm{L}^2}

\def\cinf{C^{\infty}}
\def\coinf{C_{0}^{\infty}}

\def\ft{\tilde{f}}

\def\ess{{\rm ess}}

\def\supp{{\rm supp\,}}

\def\ext{{\rm ext}}

\def\fld{\rightarrow}

\def\modk{|k|}

\def\dz{dz\wedge d\overline{z}}

\def\ball{\rm B}

\def\HN{H}
\def\KN{K_N}

\def\tauexp{E^0_{N-N'}+E_{N'}}
\def\DR{D_R}

\def\xf{ \mathrm{x}}
\def\xb{\mathit{x}}
\def\X{ X}
\def\d{{\rm d}}

\def\inter{\rm{int}}

\def\vmu{v_{\mu}}
\def\vmuj{v_{\mu,j}}
\def\vren{v^\ren}
\def\vmuren{v_\mu^{\rm ren}}

\def\Hint{\H_{\inter}}

\def\ommu{\om_\mu}
\def\dgommu{\dg(\ommu)}

\def\KNren{K^{{\rm ren}}_{N}}
\def\KNmuren{K^{{\rm ren}}_{N,\,\mu}}

\def\HNmu{H_{\mu}}
\def\HNren{H^{\rm ren}}
\def\HNmuren{H^{{\rm ren}}_{\mu}}
\def\HNmuoms{H_{\,\mu}|_{\Hint\otimes\C_\mu}}

\def\HNcmu{\hat{H}_{\mu}}

\def\HNcmuoms{\hat{H}_{\mu}|_{\Hint\otimes\C_\mu}}

\def\uqp{\frac{1}{4\pi}}

\def\sumja{\sum_{j\in a}}
\def\sumjna{\sum_{j\notin a}}

\def\sumiuN{\sum_{i=1}^{N}}
\def\sumjuN{\sum_{j=1}^{N}}

\def\joarp{j_{0,\,a,\,P}}
\def\jinfarp{j_{\infty,\,a,\,P}}
\def\jarp{j_{a,\,P}}
\def\jorp{j_{0,\,a,\,P}}
\def\jinrp{j_{\infty,\,a,\,P}}

\def\jRdue{j^2_R}
\def \chiRp{\chi_{P}}

\def\ash{a(h)}

\def\ac{a^*}
\def\ad{a^\sharp}
\def\adsh{a^\sharp(h)}

\def\ak{a(k)}

\def\gca{\check{\Gamma}^*}
\def\gc{\check{\Gamma}}
\def\dg{\d\Gamma}

\def\dgext{\d\Gamma^{\ext}(\ommu)}
\def\dgmext{\d\Gamma^{\ext}(\ommu)}

\def\phivm{\Phi(\vmu)}

\def\phivaext{\Phi_a^{\ext}(\vmu)}
\def\phivamext{\Phi_{a}^{\ext}(\vmu)}

\def\Hextsp{\mathcal{H}^{\ext}}

\def\Hcamu{\hat{H}_{a,\,\mu}}

\def\Hamext{\hat{H}_{a,\,\mu}^{\ext}}
\def\Hext{H^{\ext}}

\def\EN{E}

\def\ENmu{E_{\,\mu}}

\def\gammash{\Gamma(\frak h)}
\def\gammashuno{\Gamma(\frak h_1)}
\def\gammashdue{\Gamma(\frak h_2)}

\def\onek{\one_\cK}
\def\onegh{\one_{\gammash}}

\def\Far{F_{a,R\,}}

\def\lap{\triangle}

\def\lapxi{\triangle_{\xf_i}}
\def\grad{\nabla}

\def\diesis{\sharp}

\def\vren{v^{{\rm ren}}}

\def\schrod{ Schr\"{o}dinger }

\def\gll{binding condition }
\def\sf{ground state }

\def\ie{{\it i.e. }}

\def\qed{\hspace{\stretch{2}}$\Box$}

\def\fermions{fermions }
\def\fermion{fermion }

\def\boson{boson }

\newcommand{\bdisp}{\begin{displaymath}}
\newcommand{\eedisp}{\end{displaymath} \noindent}
\newcommand{\beq}{\begin{equation}}
\newcommand{\eeq}{\end{equation}}
\newcommand{\bet}{\begin{teo}}
\newcommand{\eet}{\end{teo}}
\newcommand{\bel}{\begin{Lemma}}
\newcommand{\eel}{\end{Lemma}}
\newcommand{\bep}{\begin{prop}}
\newcommand{\eep}{\end{prop}}
\newcommand{\bec}{\begin{Corollary}}
\newcommand{\eec}{\end{Corollary}}

\newcommand{\bear}[1]{\begin{array}{#1}}
\newcommand{\ear}{\end{array}}

\newcommand{\brmk}{\begin{rmk}}
\newcommand{\ermk}{\end{rmk}}

\begin{document}
\title{Existence and non existence of a ground state for the massless Nelson model under binding condition}
\author{Annalisa Panati\\
D\'{e}partement de Math\'{e}matiques\\
Universit\'{e} Paris Sud\\
91405 Orsay Cedex, France\\
\textit{annalisa.panati@math.u-psud.fr}}

 \date{}  \maketitle

\begin{abstract}
We consider a model describing $N$ non-relativistic particles
coupled to a massless quantum scalar field, called \emph{Nelson
model}, under a binding condition on the external potential. We
prove that this model does not admit ground state in the Fock
representation of the canonical commutation relations, but it does
in another not unitarily equivalent coherent representation.
Remark that the binding condition is satisfied for small values of
the coupling constant.
\end{abstract}


\section{Introduction}

When considering a non-relativistic atom coupled to a quantized
radiation field,
 it is natural to require that the model predicts  the existence of a ground state.
If the field is massive, this usually follows from the fact that
the bottom of the spectrum is an isolated point. On the other
hand, in the massless case the bottom of the spectrum lies in the continuum.\\
 For the standard model of non-relativistic Quantum Electrodynamics (often called Pauli-Fierz
 model) with $N$-body Coulomb interactions, the existence of a ground state in the massless case
 was first established by Bach, Fr\"{o}hlich and Sigal in \cite{bfs}
for sufficiently small values of some parameters in the theory.
Subsequently,
  Griesemer, Lieb and Loss proved  in \cite{gll}
and \cite{grie} that a ground state exists for all values of the
parameters under the following \emph{binding condition}. Let us
call $E_N$ the bottom of the spectrum of the $N$-particle
Hamiltonian with external potential $V$ and $E^0_N$ its
translation invariant part (i.e. $V$ is removed). Then the
\emph{binding condition} is
\[
 E_{N}< \tauexp \quad \textrm{for all}\quad N'<N.\qquad (B)
\]

 If the field is neglected i.e. in the framework of usual Schr\"{o}dinger operators, this condition
 is equivalent to $E_N<E_{N-1}$ since $E^0_N= N E_1^0=0$, and it
is satisfied for $N$-body Coulomb interactions if $N<Z+1$, where
$Z$ the charge of the nucleus, as proved long ago by Zhislin in
\cite{Zhislin}. In \cite{barbaroux}, Barbaroux, Chen and Vulgater
showed  that \emph{(B)} is also satisfied for the standard model
of non- relativistic QED for $N=2$, $N<Z+1$. Finally, in \cite{ll}
Lieb and Loss completed the picture by proving the statement for
any $N$ provided $N<Z+1$.\\

A natural question to ask is whether the infrared behaviour of
other (simplified) non-relativistic QED models is the
same.\newline In this paper we consider a model describing $N$
scalar non-relativistic particles (fermions) coupled to a scalar
Bose field. This is usually called the \emph{Nelson model}.
 The Hamiltonian for $N$ particles is given by

\beq \label{defH} H_{N}=\KN\otimes\one+\one\otimes
\d\Gamma(\omega)+\lambda\Phi(v). \eeq

\noindent Here $\KN$ is a Schr\"{o}dinger operator describing the
dynamics of the particles, $\lambda$ is a coupling constant,
\[
\d\Gamma(\omega):=\int |k| a^*(k)a(k) \d k,
\]
and
\[
\Phi(v):=\frac{1}{\sqrt 2}\sum_{j=1}^N\int \frac{ e^{- \ic k
\xf_j}}{|k|^{\12}} \rho(k) a^*(k)\d k +h.c,
\]
where $a^*(k), a(k)$ are the usual creation and annihilation
operators, $\rho$ an ultraviolet cutoff function.
 These objects
will be described more precisely in the next section.\\
\indent The Nelson model belongs to a class of Hamiltonian, often
called
 \emph{abstract Pauli-Fierz
hamiltonians}, which includes the so called
 \emph{generalized spin-boson models}, and
 for which the problem of the ground state has been studied
in recent years
 (see for instance \cite{ah2},\cite{ahh}, \cite{g},\cite{ggm} and references therein).
 In the case of a \emph{confining} external potential,
it is known that the Nelson model does not admit a ground state in
the Fock representation of the canonical commutation relations
(CCR) due to, heuristically speaking, too many soft photons
(\cite{g}, \cite{spohn}). Nevertheless, it is possible to find
another representation of the CCR where the ground state exists as
done by Arai in \cite{arai}. This representation is not unitarily
equivalent to the Fock one (\cite{arai}). This is called
\emph{infrared catastrophe}. The infrared problem also appears in
scattering theory. This was first studied by Fr\"{o}hlich in
\cite{fro}, and more recently by Pizzo in \cite{pizzo2} and Chen
\cite{chen}.\newline \noindent Here we consider $N$-body
interactions, more precisely we take:

\beq \label{def kn}\KN= \sum_{i=1}^N  -\frac{1}{2}\lapxi+
V(X)+I(X)\eeq where $V$, $I$ satisfy the following:
 \[\label{eq:2corps}
\left. \begin{array}{lll}
(i) & V(X)=\sum_{j=1}^{N}v(\xf_j)\quad  I(X)=\sum_{i<j}^{N}u(\xf_i-\xf_j) \quad\quad v(\xf), u(\xf): \rrt\fld\rr  ,\\[2mm]
(ii)& v=v_{sing}+v_{reg}, u=u_{sing}+u_{reg}, \textit{ where
}v_{sing} \textit{ and }
u_{sing} \textit{  have}\\[1mm]
  &  \textit{compact support,} \quad v_{reg}\textit{ and } u_{reg} \textit{  are continuous, } \\[2mm]
(iii) & v_{reg}(\xf)=O(|\xf|^{-\eps_1}) \quad\textit{ for }|\xf|\fld \infty \qquad  where \quad \eps_1>0,\\
& u_{reg}(\xf)=O(|\xf|^{-\eps_2}) \quad \textit{ for } |\xf|\fld \infty \qquad  where\quad \eps_2>0,\\[2mm]
(iv) & \textit{$v$ and $u$ are $-\lap$ bounded with relative bound
zero}.
\end{array}\right\}\textit{    (I)} \]

We will also consider the Nelson model in another representation
of the CCR, the same used in \cite{arai}, obtaining a new
Hamiltonian denoted
by $H_N^{\rm ren}$, described in the subsection \ref{subsec nelson coherent}.\\

 We prove the following:

\bet \label{main fock} Assume the \gll (B) and the hypothesis (I)
on the potentials. Then $H_{N}$ has no ground state. \eet

\bet \label{main arai}Assume the \gll (B) and the hypothesis (I)
on the potentials. Then $ H_N^{{\rm ren}}$ admits a ground state.
\eet

Moreover, if $v=-\frac{Z}{|\xf|}$ and $u(\xf)=\frac{1}{|\xf|}$,
$(B)$ is clearly satisfied for $\lambda$ small enough as
explained in Proposition \ref{binding small l}.\\

 A key point in the proofs is to guarantee that any candidate to
be a ground state must be localized in the fermion variables. In
the confined case this property follows easily from the
compactness of $(\KN+\ic)^{-1}$, while in our case it requires
some work. In particular, if $N>1$, one has to deal with photon
localization which greatly complicates the proofs. This problem
however, was already solved for the more involved standard model
in \cite{gll} and \cite{grie}, and the same proof does apply to
our case. Here we only take care to write down explicitly that the
estimates we obtain are uniform in the infrared cutoff parameter
$\mu$, which was, since the proof was split over two papers,
somewhat left to the reader.\newline
 Once the localization property is guaranteed, we can adapt techniques in \cite{g} to
  both  existence and non-existence of the ground state. Once
  again the proofs in \cite{g} make large use of the compactness of
  $(\KN+\ic)^{-1}$, which does not hold  anymore, but we can circumvent this
  difficulty using fermion localization in a more direct way.

In the one-particle case, the same problem has already been
approached in \cite{hiro} by Hirokawa (non-existence in Fock
representation) and by Sasaki in \cite{preprint} (existence in
another representation).\\

 The paper is organized as follows. In
section \ref{sec definitions} we introduce precisely the objects.
 In section \ref{sec binding} we make some useful remarks about the binding condition. Section
\ref{sec exp decay} is devoted to exponential decay. Finally
sections \ref{sec main fock} and \ref{sec main arai} are devoted
respectively to the proofs of
Theorem \ref{main fock} and Theorem \ref{main arai}.\\

\textbf{Acknowledgements}: I'm grateful to Christian G\'{e}rard
for introducing me to this problem and for many useful
discussions.

\section{Definition and basic constructions}
\label{sec definitions}
\subsection{Notation}
 We shall use the following notation:\\

 \textbf{Definition.} Let $f$ be a function in $\coinf(\rr^d)$,
  we denote by $f_R$ the operator of multiplication by $f(\frac{\xf}{R})$ in $\l2(\rr^d,\d \xf)$.\\

\textbf{Definition.} Let $A:\rr \ni R\mapsto A_R$ where $A_R$ is a
self-adjoint operator on a Hilbert space $\H$ and let $B$ be a
self-adjoint operator, $B\geq 0$. We say that $A_R=O(R^n)B$ if for
$R>>1$
$\D(|A_R|^{\12})\supset\D(B^{\12})$ and $\pm A_R\leq C(R)B$ where $C(R)=O(R^n)$. We say that $A_R=O(R^n)$ if $A_R=O(R^n)\one$.\\

If $A,B$ are two operators on a Hilbert space, we set
$\textrm{ad}_A B:=[A,B]$. The precise meaning of $[A,B]$ will be
either specified or clear from the context.\\

\textbf{Definition.}  Let $A$ be an operator on a Hilbert space
$\H_1$  and $B$ an operator on a Hilbert space $\H_2\ot\H_1$. We
introduce $T:\H_2\ot\H_1\ot\H_1\fld\H_1\ot\H_2\ot\H_1$, $T(\psi\ot
u\ot v):=u\ot \psi\ot v$, and we define \emph{twisted tensor
product}
 $A\otc B$ as $A\otc B:= T^{-1}(A\ot B)T$.\\

\textbf{Definition.} Let $\H$ be a separable Hilbert space. We say
$T\in
 L^2(\rr^d, \d x; \B(\H))$ if $T: \rr^d \ni x \mapsto T(x) \in \B(\H)$
 is a weakly  measurable function such that
 \[
\norme{T}_{ L^2(\rr^d, \d x; \B(\H))}:=\left(\int
\norme{T(x)}_{B(\H)}^2 \d x\right)^{\12} < \infty.
 \]

\subsection{Fock and coherent representations}

Here we describe some well known facts about bosonic Fock spaces
and coherent representation of CCR (for more details we refer the
reader, for instance, to \cite{dg} and \cite{ginfr}).

\subsubsection{Bosonic space and creation/annihilation operators }
 Let $\ch$ be a Hilbert space. The \emph{bosonic Fock} space over $\ch$ is the
direct sum $ \gammash:=\bigoplus_{n=0}^{\infty}\nparts$ where
$\nparts$ denote the symmetric n-th tensor power of $\ch$.  The
\emph{number operator} $\cN$ is defined as $ \cN|_{\nparts}
=n\one.$
\noindent If $h\in \ch$, we define the \emph{creation operator}
$\ac(h)$ and the \emph{annihilation operator} $\ash$ by setting,
for $u\in\bigotimes_{\rm s}^{n}\ch$,
\[
\bear{l} \ac (h) u:=\sqrt{n+1}\; u\otimes_{\rm s} h,
\\[2mm]
a(h) u:=\sqrt{n} (h|\,u. \ear
\]
\noindent By $a^{\diesis}(h)$ we mean both $a^*(h)$ and $a(h)$.\\
\noindent If the one-particle space is $ \ch=\l2(\rr^d, \d k)$,
then we can define the expressions $a(k)$, $a^*(k)$ by:
\[
\bear{l} a(h)=:\int a(k)\bar h(k)\d k
\\[2mm]
a^*(h)=:\int a^*(k)h(k)\d k. \ear
\]
\noindent We define the \emph{Segal field operators}:
\[
\Phi(h):=\frac{1}{\sqrt 2}(a^*(h)+a(h)).
\]

\noindent Let  $\Om\in \bigotimes^0 \ch $ denote the \emph{vacuum
vector}.
 \noindent There exists a large class of representations
of the CCR, called \emph{$g$-coherent representations}, which are
constructed  by defining the new creation/annihilation operators
$a^*_g$/$a_g$ acting on $\Gamma(\ch)$ as follows: let $\ch_0$ be a
dense subspace of $\ch$ and $g\in\ch_0'$ (the dual of $\ch_0$);
then we define: \beq \label{eq def ag} \bear{l}
a^*_g(h):=a^*(h)+\frac{1}{\sqrt
2}<g,h>\\[2mm]
a_g(h):=a(h)+\frac{1}{\sqrt 2}\overline{<g,h>}, \ear
\label{ag}\eeq

\noindent where $<\;,\;>$ is the duality bracket.\\

\noindent The following fact is well known  (see for instance
\cite[Theorem 3.2]{ginfr}):
 \bep \label{unitequivgrp}
 \renewcommand{\labelenumi}{(\roman{enumi})} \begin{enumerate}
    \item if $g\in\ch$, ( \ref{eq def ag}) can be rewritten as
    $ a_g^{\diesis}(h)=e^{\Phi(-\ic g)}a^{\diesis}(h)e^{\Phi(\ic
    g)}$,
    \item if $g \notin \ch$, there exists no unitary operator $U$ such that
    $a_g^{\diesis}(h)=U^*a^{\diesis}(h)U$.\\
 \end{enumerate}
 In other words a $g$-coherent
representation is unitarily equivalent to the Fock representation
if and only if $g\in\ch$.\eep

\subsubsection{The operator $\dg$}

If $b$ is an operator on $\ch$, we define the operator
$\dg(b):\gammash\fld\gammash$ by
\[\dg(b)|_{\nparts}:=\sum_{i=1}^{n}
\underbrace{\one\otimes\cdots\otimes\one}_{i-1}\otimes b
\otimes\underbrace{\one\otimes\cdots\otimes\one}_{n-i}.
\]

\noindent If $g\in\ch$, we define $\dg_g(b):=e^{\Phi(-\ic g)}\dg
(b) e^{\Phi(\ic g)}$ and one can compute that \beq
\label{dgchange} \dg_g(b)=\dg(b)+\phi(bg)+\frac{1}{2}(g,bg), \eeq
provided $b\geq0$ and $g\in D(b^{1\slash 2})$. If $\ch_0\subset
D(b^{1\slash 2})$ and $b^{1\slash 2}:\ch_0\fld\ch_0$ then by
duality $b^{1\slash 2}:\ch'_0\fld\ch'_0$. If $g\in
\ch'_0\setminus\ch$  and $b^{1\slash 2}g\in\ch$, we can make sense
of the expression  in the right hand side of (\ref{dgchange}) and
define $\dg_g(b)$ in the same way.

\subsubsection{The operators $\Gamma$ and $\gc$}

Let $\ch_i$, $i=1,2$ be two Hilbert spaces. If $q\in
B(\ch_1,\ch_2)$, we define the operator
$\Gamma(q):\gammashuno\fld\gammashdue$
\[
\Gamma(q)|_{\npartsuno}:= q\otimes\cdots\otimes q.
\]

\noindent  Let $j_1$,$j_2 \in B(\ch)$. We denote by $j=(j_1,j_2)$
the operator $ j:\ch\fld\ch\oplus\ch$ defined by $j h:=(j_1 h, j_2
h)$.
  We define the operator
$\gc(j):\G(\ch)\fld\G(\ch\otimes\ch)$ as:
 \[
 \gc(j):=U \G(j),\]
 where
$U:\G(\ch\oplus\ch)\fld\G(\ch)\otimes\G(\ch)$ is the
\emph{exponential map} defined by
 \beq \label{exp map}
U\Om=\Om \otimes \Om \qquad U\ad(h_1\oplus h_2)=(\ad(
h_1)\otimes\one+\one \otimes \ad( h_2))U, \quad h_i \in \ch.\eeq
\noindent Assume $j$ is isometric \ie  $j_1^*j_1+j_2^*j_2=1$, then
$\gca(j)\gc(j)=1$.
 Moreover, if $j_i=j_i^*$, $i=1,2$, then using (\ref{exp map}) one can easily check the following:
\begin{eqnarray}
   && \gc(j)\adsh=(\ad(j_1 h)\otimes\one+\one\otimes\ad(j_2
h))\gc(j), \label{iso1}\\[2mm]
   && \dg(b)=\gca(j)(\dg(b)\otimes \one +\one
\otimes\dg(b))\gc(j)+\frac{1}{2}\dg(ad^2_{j_1} b+ad^2_{j_2} b),
\qquad \qquad \label{iso2}
\end{eqnarray}

\noindent where $b$ is an operator on $\ch$. \\
\noindent We define the operator $\cN^{{\rm \ext}}:\gammash \ot
\gammash\fld\gammash \ot \gammash$ by
\[
\cN^{{\rm \ext}}:= \cN\ot \one + \one \ot\cN.
\]

\label{nelsonmodel}\subsubsection{The Nelson Model} \label{subsec
nelson}
 The Hilbert space
$\mathcal{H}$ is $\H: =\cK\otimes\Gamma(\ch)$
 where $\K$ is the $N$-particle space
$ \cK:=\atensorone{j}{N}\l2(\mathbb{R}^3,\d\xf_j)$ (the spin is
neglected but the Fermi statistics is kept) and $\Gamma(\ch)$ is
the Fock space with $\ch=\l2(\mathbb{R}^3,\d k)$. To avoid
confusion we denote the \fermion position by $\xf\in\rrt$, the
\boson position by $\xb\in \rrt$, $\xb:=\ic\grad_k$ and the
position of the system of $N$-\fermions by $\X\in\rrtn$,
$\X=(\xf_1,\dots,\xf_N)$.

\noindent The Hamiltonian $H_{N}$ is given by

\beq \label{defH} H_{N}=\KN\otimes\one+\one\otimes
\d\Gamma(\omega)+\lambda\Phi(v). \eeq

\noindent where $\lambda$ is a coupling constant. The operator
$\KN$ is given by (\ref{def kn}) and we assume hypothesis
\emph{(I)} given in the introduction.

\noindent  The operator $\om$ is the operator of multiplication by
the function $\om(k)$. For the sake of simplicity we consider only
the
physical case $\om(k)=\modk$.\\
\noindent The operator $\Phi(v)$ is given by
\[
\Phi(v)=\frac{1}{\sqrt 2}\int v^*(k)a(k)+ v(k) a^*(k)\d k,
\]
where $v:\rrt\fld\BK$ is defined by \[v(k):=\sumjuN \frac{e^{-\ic
k \xf_j}}{\om(k)^{\12}}\rho(k)\] where
$\rho\in\coinf(\ball(0,\La))$ with $\rho(-k)=\bar{\rho}(k)$ or
equivalently $\check{\rho}$ real.

\noindent It is well known that this Hamiltonian is well defined
and bounded from below (see for example \cite{g}).\\

\textbf{Notation:} for simplicity we will drop the dependence on
$N$ everywhere, unless it needs to be specified.

\subsubsection{Infrared cutoff Hamiltonians} We will need
\emph{infrared cutoff Hamiltonians} $\HNmu$ for $\mu>0$. We define

\beq \label{defHNmu} \HNmu:=K\otimes\one+\one\otimes
\d\Gamma(\omega)+\lambda\Phi(\vmu), \eeq

\noindent where $\vmu(k):=\chi_\mu(k) v(k)$  with
$\chi_\mu(k):=\chi(\frac{k}{\mu})$ and $\chi \in \cinf(\rrt)$ such
that $\chi\equiv 1$ for $|k|>2$ and $\chi\equiv 0$ for $|k|<1$, $\chi_\mu(-k)=\bar{\chi}(k)$.\\
\noindent
 We will also need another cutoff Hamiltonian $\HNcmu:\H\fld\H$
defined by: \beq \label{defHNcmu} \HNcmu:=K\otimes\one+\one\otimes
\dgommu+\lambda\Phi(\vmu),
 \eeq
where $\ommu\in \cinf(\rrt)$, $\ommu(k):=\mu$
$\om_1(\frac{k}{\mu}$) and
 $\om_1(k)=\om_1(|k|)$ is a smooth function, increasing with respect to $|k|$, equal to $ \om$ on $\{1<|k|\}$ and
 equal to $1-\delta$ on $\{\modk<1-\delta\}$,  $\delta<<1$. Note
 that $\ommu\geq\tilde{\mu}$ where $\tilde{\mu}=(1-\delta)\mu$.

\subsubsection {Nelson model in a coherent representation}
\label{subsec nelson coherent} We consider the Nelson model in a
$g$-coherent representation, the same originally considered by
Arai in \cite{arai}. \\

Choosing $g=-\lambda N\frac{\rho(k)}{\om(k)^{\32}}$, and using
(\ref{dgchange}) and (\ref{eq def ag}), the new Hamiltonian
becomes

\beq \label{defHren} H_N^{{\rm ren}}=\KNren\otimes\one +\one
\otimes\dg(\om)+\lambda\phi(\vren) \eeq
 where \[ \vren=v-\om g,\qquad \KNren =\KN+W(X),\]

\noindent with \[W(X)=-\lambda^2 N\sumiuN
w(\xf_i)+\lambda^2\frac{N^2}{2}w(0),\qquad w(\xf)=\int_{\rrt}
e^{\ic k\xf}\frac{|\rho|^2(k)}{\om^2(k)} \; \d k.
\]

\noindent Note that $w(\xf)=O(|\xf|^{-1})$ as  $|\xf|\fld \infty$
and that $w(\xf)=\uqp \check{\rho} *\check{\rho} *
\frac{1}{|\xf|}$where
 $\check{\rho}$ is the inverse Fourier transform of $\rho$.\\
For $\mu>0$ we use $ g_\mu=-\lambda
N\frac{\rho(k)}{\om(k)^{\32}}\chi_{\mu}(k)$ and we obtain

\beq \label{defHmuren} H_{N,\,\mu}^{{\rm ren}}=\KNmuren\otimes\one
+\one \otimes\dg(\om)+\lambda\Phi(\vmuren) \eeq with \[
\vmuren=\vmu-\om g_\mu,\]and $\KNmuren$ is the same as before
replacing  $w$ by $w_{\mu}$ with
\[w_\mu(\xf):=\int_{\rr^3} e^{\ic k\xf}\frac{|\rho|^2(k)}{\om^2(k)}|\chi_\mu|^2(k) \d k,\]
and $W$ by $W_{\mu}$ consequently.

\brmk   Because of Proposition \ref{unitequivgrp}, $\HNmu$ and
$\HNmuren$ are unitarily equi\-va\-lent while $H$ and $\HNren$ are
not.
 \ermk

 \section{Binding condition}
\label{sec binding}
 Let $A_N$ be a family of self-adjoint operators
on $\cH$ depending on $N \in \N$ and $A^0_N$ the corresponding
family of their translation invariant part. Assume that all the
operators are bounded from below. For $A$ a self-adjoint operator
let us denote $E(A):= \inf \sigma(A)$. Then we can define the
\emph{ionization threshold energy} of $A_N$ as
\[
\tau(A_N):= \inf_{0<N'\leq N}\{E(A^0_{N'})+E(A_{N-N'})\}.
\]
The binding condition then is
\[
(B)\qquad \quad E(A_N)<\tau(A_N).
\]
From a physical point of view this is a minimal condition to
require on the external potential to be binding. From a
mathematical point of view another energy can be considered. This
is the energy below which exponential decay can be proved in a
quite general way, as it was done in \cite{grie}, so we can call
it \emph{localization energy} $\Sigma(A)$. Let us define
\[\DR:=\{\psi \in \cH\,|\, \psi(X)=0 \mbox{ if  } |X|<R\}.\]
 We define
\beq \label{defsiR} \Si_R(A):=\inf_{\psi\in \DR \cap D(A),\,
||\psi||=1}(\psi,A\psi),\eeq and \beq
\label{defsi2}\Si(A):=\lim_{R\fld\infty} \Si_R (A). \eeq In
\cite{grie} it is also proved that for the standard model of
non-relativistic QED the two energies are the same; this is also
true in our case as we will explain in subsection \ref{exp
decay}.\newline

A key observation in the proof of exponential decay is that both
$\HNmu$ and $\HNcmu$ have the same ionization and localization
energy. This is stated in the following Lemma. \bel For every
$\mu>0$, $\Si(\HNmu)=\Si(\HNcmu)$ and $\tau(\HNmu)=\tau(\HNcmu)$.
\label{rmksitau}
 \eel
\proof Set  $\ch_\mu:=\l2(\{ |k|\!\leq \!\mu \})$, $\C_\mu:=\ot^0
\ch_\mu$, $\Hint:=\cK\otimes\Gamma(\ch^\perp_\mu)$. Since
$\ch=\ch_\mu \oplus \ch_\mu^\perp$, we can identify $\cH$ and
$\Hint\ot \Gamma(\ch_\mu)$ by exponential map. One can observe $
\inf \si(\HNmu)=\inf\si(\HNmuoms)$, $ \inf
\si(\HNcmu)=\inf\si(\HNcmuoms)$. Since
$\HNmuoms=\HNcmuoms$, the lemma follows. \qed\\

Thanks to the above Lemma, we can introduce the following
notation:

\[\bear{l} \Si_{R,\, \mu}:=\left\{\bear{ll}
\Si_R(\HNcmu)=\Si_R(\HNmu) \quad& \textit{for } \mu>0,\\[1mm]
\Si_R(\HN) & \textit{for } \mu=0,\ear\right.  \qquad
\tau_\mu:=\left\{\bear{ll}
\tau(\HNcmu)=\tau(\HNmu) \quad& \textit{for } \mu>0,\\[1mm]
\tau(\HN) & \textit{for } \mu=0,\ear\right.\\
\\
\Si_\mu:=\lim_{R\fld\infty}\Si_{R,\, \mu}.\ear
\]
\begin{rmk}
Since $\HNmu$ and $\HNmuren$ converge in the norm resolvent sense
to $\HN$ and $\HNren$ respectively (see \cite[Lemma A.2]{g}), and
$\HNmu$ and $\HNmuren$ are unitarily equivalent by Proposition
\ref{unitequivgrp}. Hence $E(\HNcmu) = E(\HNmu)= E(\HNmuren)$ and
$ E(H)= E(\HNren)$.
\end{rmk}

Thanks to the above remark we can give the following
definition.\newline \noindent \textbf{Definition. }We define
\[\bear{l}
\ENmu:= E(\HNcmu) = E(\HNmu)= E(\HNmuren),\\[2mm]
E:= E(H)= E(\HNren).\\[2mm]\ear
\]

It is easy to prove that if the binding condition is satisfied
when neglecting the field, then it still holds for $\lambda$ small
enough.
 \bep Assume $E(K_N)<\tau(K_N)$.
Then $E(H_N)<\tau(H_N)$  for $\lambda$ small enough.\label{binding
small l} \eep \proof Set $H_0:=\KN\ot \one+\one\ot \d
\Gamma(\omega)$ and $H_\lambda:=\HN$. Since $\Phi(v)$ is
$H_0$-bounded as operator (see \cite[Lemma 2]{nelson}) then
clearly $H_\lambda\fld H_0$ as $\lambda\fld 0$ in the norm
resolvent sense, which implies $E(H_\lambda)\fld E(H_0)$. Now
$E(H_0)=E(K_N)$ and $\tau(H_0)=\tau(K_N)$. \qed

\begin{rmk}
If $v(\xf)=-\frac{Z}{|\xf|}$ and $u(\xf)=\frac{1}{|\xf|}$, it is
well known that $E(K_N)< E(K_{N-1})$ if $N<Z+1$ ( see
\cite{Zhislin}).
\end{rmk}

\noindent It is also possible to prove that at least one particle
must be bounded. This proves \emph{(B)} for $N=1$. \bep[binding of
at least one particle] \label{binding1} Assume (I) holds and
that the operator $-\frac{1}{2}\lap+ v$ admits an eigenvalue of energy $-e_0<0$. Then $E(H_1)< \tau(H_1)$.\\
 If in addition $v(\xf)\leq 0 $ for all $\xf$, then $E(H_N)\leq E(H^0_N)-e_0$. \eep

\noindent \proof  The proof is  the same as in  \cite[Theorem
3.1]{gll}, and is therefore omitted.\qed\\

\noindent We remark that binding without mass implies binding with
mass.
 \label{bindingmass}\bep Assume $E
< \tau_0$. Then $E_\mu < \tau_\mu$ for $\mu$ small enough.\eep
\proof As $\mu\fld 0$, $\HNmu$  converges in the norm resolvent
sense to $\HN$ (see \cite[Lemma A.2]{g}). Hence as $\mu\fld 0$,
$E_\mu$ converges to $E$ and $\tau_\mu$ converges to $\tau_0$.

\section{Exponential decay}
\label{sec exp decay}
 \subsection{Localization Lemma}

Here we state a key Lemma about \boson localization needed in the
next subsection to prove exponential decay.
 \brmk The
next proposition is true for $\HNcmu$ but not for $\HNmu$. This is
one of the reason why the infrared regularization $\HNcmu$ was
introduced. \ermk
 \bel \label{localiz cor} Let $\HNcmu$ be the Hamiltonian defined in
(\ref{defH}). Then \[ \HNcmu\geq\tau_\mu-f(\mu)
o\left(R^{0}\right)(\HNcmu+C) \hbox{ on } \DR,
\]
where $f(\mu):= \frac{\ln^{\12} \mu}{\mu}$ and $\DR:= \{\psi \in
\cH\,|\, \psi(X)=0 \mbox{ if  } |X|<R\}$. \eel

\proof The proof
 is the same (simplified) as in \cite[Corollary A.2]{gll} and
 \cite[Theorem 9]{grie}. Only the dependence on the
error on the infrared parameter is different, and the estimate
needed in our case is given by the next Lemma. Notice in
\cite{gll} this dependence was not explicitly considered and it is
not uniform in $\mu$ as stated. This gap left in the proof was
filled in \cite{grie}. Since their proof is long and the reader
could get lost, we provide a sketch of the proof in our case in
Appendix A. \qed

 \bel
\label{Lemma pseudodiff2} Fix $X=(\xf_1,\ldots,\xf_2)\in \rrtn$.
Let $j\in \cinf(\rrt)$ be a function such that

\renewcommand{\labelenumi}{(\roman{enumi})}

\begin{enumerate}
    \item $0\leq j\leq 1$,
    \item $\supp j_R \in \{\xb \,|\, |\xb-\xf_j|>R\}$
    (where $j_R(\xb):= j(\frac{\xb}{R})$),
\end{enumerate}
Then $\norme{j_R(\xb) \hat{v}_\mu(\xb-\xf_j)}_{\l2(\rrt, \d
x)}=O(R^{-1})O(ln^{\12}\mu)$ uniformly in  $X$.

\eel \proof Here $\hat{v}_\mu(\xb-\xf_j)$ is
$\mathcal{F}\vmu(k,\xf_j)$, where $\mathcal{F}$ is the Fourier
transform with respect to the variable $k$. In other words we use
the unitary equivalence of the space $\ch$ and $\l2(\rrt, \d
\xb)=:\ch_\xb$ given by the Fourier transform.
 Since $ j\leq 1$ and $\supp j_R \subset \{\xb\, |\,
 |\xb-\xf_j|>R\}$, there exists a function $F\in \coinf(\rrt)$, with $F(0)=1$
 such that
\[
\norme{j_R(\xb)
\hat{v}_\mu(\xb-\xf_j)}_{\ch_{\xb}}\leq\norme{(1-F(\frac{\xb}{R}))
\hat{v}_\mu(\xb)}_{\ch_{\xb}}=\norme{(1-F(\frac{D_k}{R}))
\vmu(k,0)}_\ch,
 \]
since $\hat{v}_\mu(\xb)=\mathcal{F}\vmu(k,0)$.
By standard pseudodifferential calculus
\[
\normeh{(1-F(\frac{D_k}{R})) \vmu(k,0)}\leq
\frac{1}{R}\normeh{\grad \vmu(k,0)}=\frac{1}{R}O(\ln^{\12} \mu)
\]
as one can easily compute. \qed

\subsection{Exponential Decay} \label{exp decay}

In this subsection we  prove uniform exponential decay for states
of energy lower than the ionization energy $\tau_\mu$ for all
$\mu\geq 0$. The proof consist in two parts. First we prove
localization below $\Si_\mu$; this can be done in a more general
framework (Prop. \ref{preexpodecay}). This argument and its proof
are the same as in $\cite{grie}$, we only take care of checking
that the estimates are uniform in the infrared parameter $\mu$,
which is
left to the reader in \cite{grie}.\\
 Secondly, we prove $\Si_\mu=\tau_\mu$ for all $\mu \geq 0$. This is also as in \cite{grie}. Remark that in
\cite{grie} the two different infrared regularizations are used.
Collecting the results, one obtain exponential decay for our model
(Corollary \ref{exp decay}).

\bep[uniform exponential decay] \label{preexpodecay} Let
$\cH:=\l2(X)\ot\cH_1$, where $\cH_1$ is an auxiliary Hilbert
space. Let $H(\mu)$ for $0\leq\mu\leq 1$ be a family of
self-adjoint operators on $\cH$. Assume there exists $D$ dense in
$\cH_1$ such that $\coinf(X)\ot D$ is a core for $H(\mu)$. We
assume also
\begin{description}
\item[(i)] the IMS localization formula is satisfied \ie \[
H(\mu)= \sum_{i=1}^2 j_i H(\mu) j_i-\sum_{i=1}^2 |\grad
j_{i,R}|^2, \textit{ if
 }j_1^2(X)+j_2^2(X)=1,
\]
\item[(ii)]if $g\in \cinf(X,\rr)$ then  $e^g H(\mu) e^{-g}+e^{-g}
H(\mu) e^{g}=2H-2|\grad g|^2,$
\end{description}
as quadratic forms on $\coinf(X) \ot D$.\newline \noindent Let us
denote $\Si_R(\mu):=\Si_R(H(\mu))$, $\Si(\mu):=\Si(H(\mu))$ and
$E(\mu):= \inf \si(H(\mu))$. We suppose
\begin{description}
\item[(iii)] $ E(\mu)\fld E(0)$,\item[(iv)] $\Si_R(\mu)\fld
\Si_R(0)$ \textit{uniformly in }R,
\end{description}
as $\mu\fld 0$.\\
Fix a function $g\in \cinf(X,\rr)$ such that $|\grad g|^2\leq 1$.\\[1mm]
\noindent  If   $\;\Si(0)-E(0)= \delta_0 >0$ then for all $f\in
\coinf(\rr)$ with $\supp f \subset ]-\infty, \Si(0)[$ there exists
 $\mu_0$ and  $\beta_0$ (depending only on $f$) such that for all
$\mu \leq \mu_0$, $\beta <\beta_0$

\[\norme{e^{\beta g}f(H(\mu))}\leq C \textit{    uniformly in }\mu.\]
\eep

\proof Since $\supp f \subset ]-\infty, \Si(0)[$ we can assume
that $\supp f \subset ]-\infty, \Si(0)-\alpha]$, $\alpha>0$. Since
$E(\mu)\fld E(0)$ and $\Si(\mu)\fld \Si(0)$, there exists $\mu_0$
such that for all $\mu\leq\mu_0$

 \beq \label{tau}
|\Si(\mu)-\Si(0)|<\frac{\alpha}{4},\qquad \eeq
\[
|E(\mu)-E(0)|<\frac{\alpha}{4},\qquad
|\Si(\mu)-E(\mu)|>\delta_0-\frac{\alpha}{2}.
 \]
By  (\ref{tau}) $\supp f \subset ]-\infty, \Si(\mu)[$ for all
$\mu\leq \mu_0$.

 Let us denote
\[
    H_{R,\,\mu}:=H(\mu)+(\Si_R(\mu)-E(\mu))\chi_R
\] where $\chi$ is a smoothed characteristic function of
the unit ball. Take $j_1,j_2 \in \cinf(X)$ such that $j_1^2
+j_2^2=1$, $j_1=1$  on $\ball(0,\frac{1}{2})$ and $j_1=0$ outside
$\ball(0,1)$. By the IMS localization formula
$H_{R,\,\mu}=\sum_{i=1}^2 j_{i,R} H_{R,\,\mu}j_{i,R}-\sum_{i=1}^2
|\grad j_{i,R}|^2$.
 \noindent By the definition of $H_{R,\,\mu}$
and since $\chi_R=1 $ on $\supp j_{1,R}$

\[
\begin{array}{l}
 j_{1,R}H_{R,\,\mu} j_{1,R}=j_{1,R}(H(\mu)+\Si_R(\mu)-E(\mu))j_{1,R}\geq \Si_R(\mu)  j_{1,R}^2, \\[2mm]
 j_{2,R}H_{R,\,\mu} j_{2,R}\geq \Si_R(\mu)  j_{2,R}^2.\\
\end{array}
\]
\noindent Hence
\[
H_{R,\,\mu}\geq \Si_R(\mu)(j_1^2 +j_2^2)-\sum_{i=1}^2 |\grad
j_{i,R}|^2 \geq \Si_R(\mu)- C R^{-2},
\]
  uniformly in
 $\mu$.
For $R\geq R_0$ and $\mu\leq\mu_0$ \[ \Si_R(\mu)-
\frac{C}{R^2}\geq \Si_R(\mu)- \frac{\alpha}{4} \geq \Si(0)-
\frac{\alpha}{2}. \] If $\la \in \supp f$, then  $\la\leq
\Si(0)-\alpha=E(0)+\delta_0-\alpha$. We have
$\Si(0)-\frac{\alpha}{2}>E(0)+\delta_0-\frac{\alpha}{4}$.
Hence for $R\geq R_0$, $ f(H_{R,\mu})=0$ for $\mu\leq \mu_0$.\\
\noindent We want now to show (for any \emph{fixed}
 $R\geq R_0$)
\[ e^{\beta g}f(H(\mu))=e^{\beta g}(f(H(\mu))-f(H_{R,\mu}))    \hbox{ is bounded  uniformly in $\mu$}.\]
By Theorem \ref{almost analytic} we can write

\[
\begin{array}{rl}
  e^{\beta g}&f(H(\mu))=e^{\beta g}(f(H(\mu))-f(H_{R,\mu}))= \\[2mm]
&=\frac{1}{\ic\pi}\int_{\C} \dd{\ft}{\bar{z}}(z)e^{\beta
g}(z-H_{R,\mu})^{-1}e^{\beta g} e^{-{\beta
g}}(\Si_R(\mu)-E(\mu))\chi_{R}(z-H(\mu))^{-1}\dz.
\end{array} \]

\noindent Since $\supp \ft$ is compact, it suffices to estimate
the integrand:
\[
\begin{array}{l}
  \norme{e^{\beta g}(z-H_{R,\mu})^{-1}e^{\beta g}
e^{-{\beta g}}(\Si_{R}(\mu)-E(\mu))\chi_{R}(z-H(\mu))^{-1}}\leq \\[2mm]
  \;\;\leq \norme{e^{\beta g}(z-H_{R,\mu})^{-1}e^{-{\beta g}}}\norme{
e^{{\beta g}}\chi_{R}}_{\infty}|\Si_{R}(\mu)-E(\mu)|\norme{(z-H(\mu))^{-1}}. \\[2mm]
\end{array}
\]

\noindent We have  $\norme{e^{{\beta g}}\chi_{R}}_{\infty}\leq
e^{\beta g_R}$ where $g_R:= \sup_{\{ |X|<R\}}g(X)$, \hspace{1mm}
$\Si_{R}(\mu)-E(\mu)\leq \Si(\mu)-E(\mu)$,
$\norme{(z-\HN)^{-1}}\leq |{\Im z}|^{-1}$.  It remains to estimate
$ \norme{e^{\beta g}(z-H_{R,\,\mu})^{-1}e^{-{\beta g}}}$. First we
notice that
\[e^{\beta g}(z-H_{R,\,\mu})^{-1}e^{-{\beta g}}=(z- e^{\beta g}
H_{R,\,\mu}e^{-{\beta g}})^{-1}\] and that

\[
\begin{array}{l}
\Re  (e^{\beta g} H_{R,\,\mu}e^{-{\beta g}}-z)= H_{R,\,\mu}
-\beta^2 |\grad {g}|^2- \Re  z\\[2mm]\geq
\Si(\mu)-\frac{\alpha}{4}-\beta^2-(E(0)+\delta_0-\alpha)\\[2mm]
\geq \Si(0)-\frac{\alpha}{2}-\beta^2-(E(0)+ \delta_0-\alpha)
\geq\frac{\alpha}{2}-\beta^2.\\[2mm]
\ear
\]
This implies for $\beta^2< \frac{\alpha}{2}$ that $\norme{e^{\beta
g}(z-H_{\mu,\,R})^{-1}e^{-{\beta g}}}
\leq\frac{1}{\frac{\alpha}{2}-\beta^2}$ uniformly in $\mu$.

\noindent Collecting all the estimates we obtain
\[
 \norme{e^{\beta g} f(H(\mu))}
 \leq \frac{(\Si(0)-E(0)+\frac{\alpha}{2})
 e^{\beta g_R}}{\frac{\alpha}{2}-\beta^2}.\]

\qed

\bel \label{lemma conv Hmu}
\renewcommand{\labelenumi}{(\roman{enumi})}
\begin{enumerate}
    \item $\Si_\mu<\infty$ if and only if $\Si_0<\infty$ and in this case there exists a constant $C$ such that
     $\left| \Si_{R,\, \mu}-\Si_{R,\, 0}\right|<C\mu^{1\slash 2}
     \textrm{    uniformly in R}$,
    \item
    there exists a constant $C$  such that
     $\left| \tau_\mu-\tau_0\right|< C\mu^{1\slash 2}$.
\end{enumerate}
\eel \proof  The proof is the same as in \cite[Proposition
5]{grie} just noting that the estimates concerning \textit{(i)}
are uniform in $R$.\qed

\bep \label{simu=taumu} \label{si=tau} For every $\mu\geq 0$,
$\Si_\mu=\tau_\mu$.\eep

\proof As in \cite [Theorem 6]{grie}, using Lemma \ref{localiz
cor}.\qed

\bec \label{ins} \label{proptec} \label{cor expdecay} Assume (B)
and (I). Let
\[H(\mu):=\left\{\bear{l}\HNmu \quad or\quad \HNcmu \quad for\quad\mu>0,\\[1mm]
\HN \quad for\quad\mu=0.
 \ear \right.
\]
Then then for all $f\in \coinf(\rr)$ with $\supp f \subset
]-\infty, \tau_0[$ there exists a $\mu_0$ and $\beta_0 $
(depending only on $f$)
 such that  for all $\mu \leq \mu_0$, $\beta<\beta_0$

\[\norme{e^{\beta |X|}f(H(\mu))}\leq C \textit{    uniformly in }\mu.\]
 \eec

\proof We apply Proposition \ref{preexpodecay} with
$\cH_1=\Gamma(\ch)$. Hypothesis (i) and (ii) are clearly verified
by direct computation, hypothesis (iii)is true since both $\HNmu$
and $\HNcmu$ converges in the norm resolvent sense to $\HN$ (see
\cite[Lemma A.2]{g}) and (iv) follows from Lemma \ref{lemma conv
Hmu}.
 By \textit{(B)} $\Si_0-\EN=\delta_0>0$.
Choose $g_\eps:=\frac{\langle X\rangle}{1+\eps \langle X \rangle}$, then $|\grad g_\eps|\leq
1$ uniformly in $\eps$ and $\sup_{\{|X|<R\}} g_\eps \leq R$
uniformly in $\eps$. Hence by Proposition \ref{preexpodecay},
there exists a $\mu_0$
 such that  for all $\mu \leq \mu_0$, $\beta\leq \beta_0$,
 $\norme{e^{\beta |X|}f(H(\mu))}\leq C$     uniformly in
$\mu$. \qed

\section{Proof of theorem \ref{main fock}}
\label{sec main fock}  We can now prove the non-existence of the
ground state
 in the Fock representation. We
use the following Lemma from \cite[Lemma 2.6]{ginfr}.

\bel \label{argomentostandard} Let $\psi\in \H$ be such that $
\int_{\rrt} \normeH{\ak \psi+h(k)\psi}^2 dk <\infty $ where $k\fld
h(k)\in \C$ is a mesurable function and $\int_{\rrt} |h(k)|^2 \d
k=\infty.$\\ Then $\psi=0$.
 \eel

 \textbf{Proof of Theorem \ref{main fock}.} Suppose there exists $\psi$ such that $\HN \psi= \EN \psi$. We
want to prove $\psi\equiv0$.
 By $\emph(B)$, there exists a function
 $f\in\mathcal{C}_{0}^\infty(\rr)$  such that  $\supp f \subset ]-\infty,\tau_0[$ and $f \equiv 1$ on an interval $[\EN,\EN+\delta]$, $\delta>0$.
By Corollary \ref{proptec},  $\vert X\vert f(\HN)$ is  bounded, so
$\vert X\vert= \vert X\vert f(\HN)\psi$ belongs to $\mathcal{H}$.\\
\indent By the pullthrough formula (as an identity on $\l2_{{\rm
loc}}(\rrt\backslash\{0\}, \d k ; \cH)$) we have
\begin{displaymath} -a(k)\psi=(\HN-\EN+\omega(k))^{-1} \sumiuN
e^{-\ic k\xf_i}\frac{\rho(k)}{\omega(k)^\frac{1}{2}}\psi.
\end{displaymath}
\noindent Writing $e^{-\ic k\xf_i}$ as $e^{-\ic
k\xf_i}=1+r(\xf_i,k)$ with $|r(k,\xf)|\leq |k||\xf|$, the former
expression becomes:

\begin{displaymath}
-a(k)\psi= m(k)\psi + h(k)\psi\end{displaymath} where
\[m(k):=\sumiuN(\HN-\EN+\omega(k))^{-1} r(k,\xf_i)
\frac{\rho(k)}{\omega(k)^\frac{1}{2}}, \qquad
h(k):=N\frac{\rho(k)}{\omega(k)^\frac{3}{2}}.\] Since $\vert
X\vert\psi \in \mathcal{H}$, clearly  $\vert \xf_i\vert \psi \in
\mathcal{H}$ and $\norme{r(k,\xf_i)\psi}\leq\modk \norme{\vert
\xf_i\vert \psi}$, which implies $m(k)\in \l2(\rrt, \d k)$.
\noindent So we have
 $\int \normeH{m(k)}^2\d k =\int \normeH{(a(k)+
h(k))\psi}^2\d k < \infty$. Since $h(k) \notin \l2(\rrt, \d k)$, then  $\psi$ must be $0$ by Lemma \ref{argomentostandard}.\qed\\

\section{Proof of theorem \ref{main arai}}
\label{sec main arai}

\subsection{Existence of ground state in the massive case}
 We adapt the proof in \cite{g}. We want to prove the
existence of a ground state for $\HNmuren$. Because of Proposition
\ref{unitequivgrp}, $\HNmuren$ is unitarily equivalent to $\HNmu$;
by Lemma \ref{equivgs}, $\HNmu$ admits a ground state if and only
if $\HNcmu$ does. We prove the existence of a ground state for
$\HNcmu$ by showing that there is a spectral gap.

\bel \label{equivgs}$\HNcmu$ admits a ground state  if and only if
$\HNmu$ admits a ground state. \eel
\proof  See \cite[Lemma 3.2 ]{g} .\qed\\

\bel\label{proHVZ}Let $\chi \in \coinf(\rr)$ be a function such
that $\supp \chi \subset ]-\infty, \tau_\mu[$. Let $j\in
\coinf(\rrt)$ be such that  $j =1$ on $B(0,1)$ and $j = 0$ outside
$B(0,2)$. Assume (I) and (B).Then the operator
$\Gamma(\jRdue)\chi(\HNcmu)$ is compact. \eel

\proof Let us denote $C:=\Gamma(\jRdue)\chi(\HNcmu)$. It suffices
to prove that $C^*C$ is  compact.
We have
\[
\bear{ll}
C^*C=\chiRp(\HNcmu) \Gamma(j_R^4)\chiRp(\HNcmu)= \\[2mm]
\; = \chiRp(\HNcmu) (-\lap_{\xf}^{\12}+ |X|+1)
(-\lap_{\xf}^{\12}+ |X|+1)^{-1}\times \\[2mm]
\; \times \;\Gamma(j_R^4)\;(\dg(\ommu)+1)^{-1} (\dg(\ommu)+1)\chi(\HNcmu).\\[2mm]
\ear
\]

The operator $\chi(\HNcmu)(-\lap_{\xf}^{\12})$ is bounded since
$D(|\HNcmu|^{\12})$=$D(|H_0|^{\12})$ where
$H_0=-\frac{1}{2}\lap_{\xf}\ot \onegh+\onek\ot \dg(\om)$. Since
$\supp \chi \subset ]-\infty, \tau_\mu[$, by Corollary \ref{ins}
also $\chi(\HNcmu)|X|$ is bounded. Hence the operator
$B_1:=\chi(\HNcmu)(-\lap_{\xf}^{\12}+ |X|+1)$  is bounded on
    $\cK\ot \gammash$.\\
Moreover $K_1:= (-\lap_{\xf}^{\12}+ |X|+1)^{-1}$ is compact on
$\cK$, $K_2:= \Gamma(j_R^4)(\dg(\ommu)+1)^{-1}$ is compact on
$\gammash$, $B_2:= (\dg(\ommu)+1)\chi(\HNcmu)$ is bounded on
$\cK\ot \gammash$.

This implies that operator $C^*C=B_1 (K_1\ot K_2)B_2$ is compact
on $\cK\ot\gammash$. \qed

\bel\label{proHVZ2}  Let $j:=(j_0,j_\infty)$ where $ j_0=1$ on
$B(0,1)$, $j_0=0$ outside $B(0,2)$ and $j_\infty$ such that
$j_0^2+j_\infty^2=1$. Let  $\Hext:=\HNcmu\ot \onegh+\onek\ot
\dg(\ommu)$. Then \[
\chi(\Hext)\gc(j_R)-\gc(j_R)\HNcmu=O(R^{-1})\]
 \eel \proof By Theorem \ref{almost analytic}
\[ \bear{l}
\chi(\Hext)\gc(j_R)-\gc(j_R)\HNcmu=\\[2mm] \quad =\frac{1}{\ic\pi}\int_{\C}
\dd{\tilde{\chi}}{\bar{z}}(z)\gc(j_R)(z-\HNcmu)^{-1}-(z-\Hext)^{-1}\gc(j_R)\dz\\[2mm]
\quad =\frac{1}{\ic\pi}\int_{\C} \dd
{\tilde{\chi}}{\bar{z}}(z)(z-\HNcmu)^{-1}(\gc(j_R)\HNcmu-\Hext\gc(j_R))(z-\Hext)^{-1}\dz.\ear
\]
Since $\supp \tilde{\chi}$ is compact it suffices to prove that $
\Hext\gc(j_R)-\gc(j_R)\HNcmu=O(R^{-1})(\cN^{\ext}+1), $ which is
equivalent to prove that $\HNcmu - \gca (j_R) \Hext \gc(j_R)=O(
    R^{-1})(\cN+1)$.\\
 \noindent We have
\[
\bear{l} \HNmu - \gca (j_R) \Hext \gc(j_R)=\dgommu-\gca(j_R)\dgmext \gc(j_R)\\[2mm]
\quad+ \lambda  \left(\phivm-\gca(j_R)\phivm\ot\onegh
\gc(j_R)\right).\\
\quad \ear\]

\noindent Now using (\ref{iso1})
\[
\bear{l} \dgommu-\gca(j)\dgmext \gc(j)=\dg (ad _{j_{0,\,R}}^2
\ommu+
ad_{j_{\infty,\,R}}^2 \ommu)\\[2mm]
\quad \leq \normeh{ad _{j_{0,\,R}}^2 \ommu+ ad_{j_{\infty,\,R}}^2
\ommu}(N+1)\leq O(R^{-1})(\cN+1) \ear\] and using (\ref{iso2})
\[
\bear{l} \phivm-\gca(j_R)\phivm\ot\onegh \gc(j_R)=\\[2mm]
\quad =\gca(j_R) \left(
\phi((j_{0,\,R}-1)\vmu)\ot\onegh+\onegh\hat{\ot}\,\phi(j_{\infty,\,R}\vmu))
\right)\gc(j_R)\\[2mm]\quad\leq
(\normeh{(j_{0,\,R}-1)\vmu}+\normeh{j_{\infty,\,R}\vmu})(\cN+1)\leq
O(R^{-1})(\cN+1)\\
\ear\] because of Lemma \ref{Lemma pseudodiff2}.  \qed

 \bet[Existence of spectral gap for $\HNcmu$] \label{massive gs} Let $\HNcmu$ be the
Hamiltonian defined in \ref{defHNcmu}. Assume (I) and (B). Then
$\si_\ess(\HNcmu)\subset[G_\mu,+\infty[$, where $G_\mu=min \{
\ENmu+\tilde{\mu}, \tau_\mu\}$ with $\tilde{\mu}=\mu(1-\delta)$,
$\delta<<1$. Consequently $\HNcmu$, $\HNmu$ and $\HNmuren$ admit a
ground state.\eet

\proof As in \cite[Theorem 4.1]{g}, using Lemma 7.2 (instead of
\cite[Lemma 4.2]{g}).
\qed\\

\subsection{Existence of a ground state in the massless case}

Let $\psi_\mu$ be a ground state for $\HNmuren$. We will prove
Theorem \ref{main arai} by showing that $\HNren$ admits a ground
state as limit of $\psi_\mu$ for $\mu\fld 0$.

As mentioned in the introduction,  the proofs in the confined case
make use of the compactness of the operator $(\KNren+\ic)^{-1}$,
which does not hold anymore. Instead, we will use the localization
in the fermion variables to control directly the behaviour as
$k\fld 0$ of $\normeH{\vren(k)\psi_\mu}$. The facts we need  are
collected in the following Lemma.

\bel We have \beq \label{vren1}\normebk{\vren(k) \langle X
\rangle^{-1}}= \normebk{\sumiuN \frac{(e^{-\ic k \xf_i}-1)\langle
X \rangle^{-1}\rho(k)}{\om(k)^{\12}}}\sim |k|^{\12}.
\label{vren2}\eeq
  \noindent Assume $(B)$ and $(I)$. Then  for all $N \in \N$,
and $\mu$ small enough \beq \label{vren3}\label{boundX}
(\psi_{\mu}, \langle X \rangle ^N\psi_{\mu})\leq C, \hbox{
uniformly in }\mu>0. \eeq Moreover  \beq \label{vren4} \int
\frac{\normeH{\vren \psi_{\mu}}^2}{\om(k)^\alpha}\d k \leq C
\hbox{    uniformly in $\mu$ for $\alpha<4$}
 \eeq
and  \beq
\label{vren5}\int\frac{\chi_\mu(k)}{\om(k)^\alpha}\normeH{\vren
\psi_{\mu}}^2 \d k =\left\{ \bear{l}O(ln \mu) \quad  if \quad
\alpha=4\\
O(\mu^{4-\alpha}) \quad if \quad \alpha<4,\ear\right. \eeq where
$\chi_\mu$ is the infrared cutoff function. \eel
 \proof
(\ref{vren1}) is obtained by direct computation, (\ref{vren3})is a
consequence of Corollary \ref{cor expdecay}. Then (\ref{vren4})
and (\ref{vren5}) follow easily by writing
$\vren(k)\psi_\mu=\vren(k)\langle X \rangle^{-1}\langle X
\rangle\psi_\mu$ and using (\ref{vren1}) and
(\ref{vren3}).\qed\\

We need some uniform bounds on $\psi_\mu$.

\bel\label{boundN}  Assume (I) and (B). Then for $\mu$ small enough
\[(\psi_{\mu},
N\psi_{\mu})\leq C \hbox{ uniformly in }\mu>0.  \] \eel

\proof By the pullthrough formula
\[\bear{ll}
(\psi_{\mu}, N\psi_{\mu})&\leq \int  \normeH{a(k) \psi_{\mu}}^2 \d
k
=\int  \normeH{(\HNmuren-\ENmu+\om(k))^{-1} \vmuren \psi_{\mu}}^2\d k\\[3mm]
&\leq \int
 \frac{1}{\om(k)^2}\normeH{\vmuren\psi_\mu}^2 \d k
\leq C\ear\]
 \noindent uniformly in $\mu$ because of
(\ref{vren4}).\qed

\bel\label{boundH0} Let $H_0:=K^{{\rm ren}}\ot\one+\one
\ot\dg(\om)$.
 Then \[ (\psi_{\mu}, H_0\psi_{\mu})\leq C
\hbox{ uniformly in }\mu>0.  \] \eel \proof As quadratic form
$\HNmuren$ is equivalent to $H_0$ uniformly in $\mu$.\qed \\

\bel \label{bound energy} Let $E:=\inf \si(\HNren)$, $E_\mu:= \inf
\si(\HNmuren)$. Assume (I) and (B). Then
\[
E-E_\mu =O(\mu).
\]
\eel\proof Let $0<\mu^\prime<\mu$. We have
\[
\bear{l} E_{\mu^\prime}-E_\mu\leq (\psi_\mu,
(\HNren_{\mu^\prime}-\HNmuren )\psi_\mu)=(\psi_\mu,
(W_{\mu^\prime}-W_\mu)\psi_\mu)+(\psi_\mu,
\Phi(\vren_{\mu^\prime}-\vmuren)\psi_\mu),\\[2mm]

E_{\mu}-E_{\mu^\prime}\leq (\psi_{\mu^\prime},
(\HNmuren-\HNren_{\mu^\prime})\psi_{\mu^\prime})=(\psi_{\mu^\prime}\,,
(W_{\mu^\prime}-W_\mu)\psi_{\mu^\prime})+(\psi_{\mu^\prime}\,,
\Phi(\vren_{\mu^\prime}-\vmuren)\psi_{\mu^\prime}).\ear
\]

\noindent Notice that  $|W_{\mu^\prime}(X)-W_\mu(X)|\leq
C(\mu^\prime-\mu)$ uniformly in $X$, hence
\[
(\psi_\mu, (W_{\mu^\prime}-W_\mu)\psi_\mu)\leq C|\mu^\prime-\mu|.
\]

\noindent Writing $\psi_\mu=\langle X \rangle^{-1}\langle X
\rangle\psi_\mu$, using Schwarz inequality and
$\norme{a(h)\psi}\leq \normeh{h}(\psi,(\cN+1)\psi)^{\12},$ we
obtain
\[\bear{l}
(\psi_\mu, \Phi(\vren_{\mu^\prime}-\vmuren)\psi_\mu)\leq C
\left(\int \normebk{(\vren_{\mu^\prime}(k)-\vmuren(k))\langle
X\rangle^{-1}}^2 \d k\right)^{\12} (\psi_\mu, (\cN+1)
\psi_\mu)^{\12} \normeH{\langle X \rangle \psi_\mu}. \ear\] The
last two terms of the right hand side product are bounded
uniformly in $\mu$ by Lemmas \ref{boundN} and (\ref{boundX}).
Hence by (\ref{vren1})
\[
(\psi_\mu, \Phi(\vren_{\mu^\prime}-\vmuren)\psi_\mu)\leq
C(\mu^\prime-\mu)^2.
\]
Estimating in the same way $E_\mu-E_{\mu^\prime}$, we obtain
$|E_{\mu}-E_{\mu^\prime}|\leq C|\mu^\prime-\mu|$. Since $E=
\lim_{\mu \fld 0}E_\mu$ the lemma follows by letting $\mu^\prime
\fld 0$.
 \qed\\

 \bep\label{prop4.4!!}
$a(k)\psi_{\mu}- (\EN-\HNren-\omega(k))^{-1}\vren(k)\psi_{\mu}\to
0$ when $\mu\fld 0$ in $L^{2}(\rrt, \d k;\cH)$. \eep

 \proof By the
pullthrough formula
\[
\begin{array}{rl}
&a(k)\psi_{\mu}- (\EN-\HNren-\omega(k))^{-1}\vren(k)\psi_{\mu}\\[2mm]
=&(\ENmu-\HNmuren-\omega(k))^{-1}\vmuren(k)\psi_{\mu}-
(\EN-\HNren-\omega(k))^{-1}\vren(k)\psi_{\mu}\\[2mm]
=&-(1-\chi_\mu)(k)(\EN-\HNren-\omega(k))^{-1}\vren(k)\psi_{\mu}\\[2mm]
\:&+(\ENmu-\EN)(\EN-\HNren-\omega(k))^{-1}(\ENmu-\HNmuren-\omega(k))^{-1}\vmuren(k)\psi_{\mu}\\[2mm]
\:&+ (\EN-\HNren-\omega(k))^{-1}(W(X)-W_{\mu}(X))
(\ENmu-\HNmuren-\omega(k))^{-1}
\vmuren(k)\psi_{\mu}\\[2mm]
&+ (\EN-\HNren-\omega(k))^{-1}(\Phi(\vren)-\Phi(\vmuren))
(\ENmu-\HNmuren-\omega(k))^{-1}
\vmuren(k)\psi_{\mu}\\[2mm]

=:& R_{\mu,1}(k)+ R_{\mu,2}(k)+ R_{\mu,3}(k)+R_{\mu,4}(k).
\end{array}
\]

Note that because of the ultraviolet cutoff, $\vren(k)$ is
compactly supported in $k$. Therefore the behaviour of the terms
for large $k$ is not relevant. First we estimate $R_{\mu,1}(k):$
\[
\|R_{\mu,1}(k)\|_{\cH}\leq\one_{\{\omega(k)\leq\,
\mu\}}(k)\frac{1}{\omega(k)}\|\vren(k)\psi_{\mu}\|_{\cH},
\]
which by  (\ref{vren4}) implies
$R_{\mu,1}\in o(\mu) \hbox{ in }L^{2}(\rrt, \d k;\cH)$.\\
\indent Now we estimate $R_{\mu,2}(k).$   By Lemma \ref{bound
energy},
 $\EN-\ENmu=O(\mu)$, then

\[
\begin{array}{l}
\normeH{R_{\mu,\,2}(k)}
\leq\frac{O(\mu)}{\om(k)^2}\normeH{\vmuren(k) \psi_\mu},\\
\end{array}
\]
hence by  (\ref{vren5}) $\norme{R_{\mu,\,2}}_{L^{2}(\rrt, \d
k;\cH)}=O(\mu \ln^{\!\12}\!\mu).$\\
The same bound holds for $R_{\mu,3}$, noticing that
$|W(X)-W_\mu(X)|\leq
O(\mu)$ uniformly in $X$.\\
\indent Finally we estimate $R_{\mu,4}$. Let $\chi\in\coinf(\rr)$
be a function such that $\supp \chi \subset]-\infty, \Si(\HN)[
\;.$ By Corollary \ref{proptec}
 and Proposition \ref{unitequivgrp}  $\chi(\HNren)\langle X \rangle$ is a bounded operator.
 Since $E\notin \supp (1-\chi)$,
 the following estimate holds for all $u\in \cH$, for $\la\!>\!0$:
\[\bear{l}
\norme{(E-H-\lambda)^{-1}u}
\leq\norme{(E-H-\lambda)^{-1}\chi(H)\langle X \rangle\langle X
\rangle^{-1}u}\\[2mm] \qquad+\norme{(E-H-\lambda)^{-1}(1-\chi(H))u} \leq
\frac{C}{\lambda}\norme{\langle X \rangle^{-1} u} + C \norme{u}.
\ear\]
 Hence
\[
\begin{array}{l}
\normeH{R_{\mu,4}(k)}  \leq  \frac{C}{\om(k)}\normeH{ \Phi(\langle
X \rangle^{-1}(\vren-\vmuren))(\ENmu-\HNmuren-\omega(k))^{-1}
\vmuren(k)\psi_{\mu}}\\[2mm]
\qquad\qquad \quad+ \normeH{
\Phi(\vren-\vmuren)(\ENmu-\HNmuren-\omega(k))^{-1}
\vmuren(k)\psi_{\mu}}.\\ \ear
\]
Since $ \norme{\phi(v)(H_0+C)^{-\12}} \leq
\left(\int\normebk{v(k)}^2\left(\frac{1}{\om(k)}+1\right)\,\d
k\right)^{\12}$ and\newline \noindent
$\norme{(H_0+C)^{-\12}(E_\mu-H_\mu - \om(k))^{-1}}\leq
\frac{C}{\om(k)}$ one obtains
$$
\bear{l}
 \normeH{R_{\mu,4}(k)}\leq \frac{C}{\om^2(k)}\left(\int
\normebk{\langle X
\rangle^{-1}(\vren-\vmuren)(k)}^2\left(\frac{1}{\om(k)}+1\right)\,\d
k\right)^{\12}  \normeH{\vmuren(k) \psi_{\mu}}\\[2mm]
\qquad\qquad\quad+\frac{C}{\om(k)}\left( \int
\normebk{(\vren-\vmuren)(k)}^2\left(\frac{1}{\om(k)}+1\right)\,\d
k\right)^{\12}\normeH{\vmuren(k) \psi_{\mu}}.\\\ear
$$

\noindent By writing $\vren-\vmuren= \vren(1-\chi_\mu)$ and by
(\ref{vren1}), one can easily check that:
\[
\left(\int \normebk{\langle X
\rangle^{-1}(\vren-\vmuren)(k)}^2\left(\frac{1}{\om(k)}+1\right)\,\d
k\right)^{\12}=O(\mu^{\32}),
\]
and
\[
\left(\int
\normebk{(\vren-\vmuren)(k)}^2\left(\frac{1}{\om(k)}+1\right)\,\d
k\right)^{\12}=O(\mu^{\12}).
\]
Then
 by (\ref{vren5}) $\norme{R_{\mu,4}}_{\l2 (\rrt, \d k;
\cH)}=o(\mu)$. \qed

\bel
\label{lemma Tk}Let us denote $T(k):= (\EN-\HNren-\omega(k))^{-1}\vren(k)\langle X\rangle^{-1}$. Then

\beq \label{mah} T(k) \textrm{  belongs to  } L^{2}(\rrt, \d k;
B(\cH)), \eeq and \beq \label{eq
Tk}\norme{T(k)-T(k+s)}_{L^{2}(\rrt, \d k; B(\cH))}\fld 0
\quad\textrm{  as }\quad s\fld 0. \eeq \eel
\begin{rmk}
Note that in general (\ref{eq Tk}) does  \emph{not} follow from
(\ref{mah})
 since $B(\cH)$ is \emph{not} a separable Banach space, but is verified for the specific element $T(k).$
\end{rmk}

\proof Set $\frak{H}:=L^{2}(\rrt, \d k; B(\cH)).$ We have
$\norme{T(k)}^2_{B(\cH)}\leq
\frac{1}{\om(k)^2}\normebk{\vmu(k)\langle X \rangle^{-1} }^2$
which is integrable by (\ref{vren1}). This prove (\ref{mah}).\\
\noindent For $0<C_1<C_2$, let us denote $K_2:=[0,\,C_1[$,
$K_2:=[C_1,\,C_2[$ and $G:=[C_2,\,\infty[$; then we can write
$\one=\one_{K_1}(|k|)+\one_{K_2}(|k|)+\one_{G}(|k|)$, so
$T(k)=\one_{K_1}(|k|)T(k)+\one_{K_2}(|k|)T(k)+\one_{G}(|k|)T(k)$.
So we can write
\[
\begin{array}{l}
T(k+s)-T(k)=\one_{K_1}(|k+s|)T(k+s)-\one_{K_1}(|k|)T(k)\\[2mm]
\,\;+
\one_{K_2}(|k+s|)T(k+s)-\one_{K_2}(|k|)T(k)+\one_{G}(|k+s|)T(k+s)-\one_{G}(|k|)T(k).
\end{array}\]

\noindent We have
\[
\begin{array}{l}
\norme{\one_{K_1}(|k+s|)T(k+s)-\one_{K_1}(|k|)T(k)}_{\frak{H}}
\leq 2 \norme{\one_{K_1}(|k|)T(k)}_{\frak{H}}\\[2mm]
\norme{\one_{G}(|k+s|)T(k+s)-\one_{G}(|k|)T(k)}_{\frak{H}} \leq 2
\norme{\one_{G}(|k|)T(k)}_{\frak{H}},
\end{array}
\]

\noindent  but on the other hand
$\norme{\one_{K_1}(|k|)T(k)}_{\frak{H}}\fld 0$   as  $C_1\fld 0$
and $\norme{\one_{G}(|k|)T(k)}_{\frak{H}}\fld 0$  as $C_2\fld
\infty$, since $T(k)\in  \frak{H}$.\\
Let us now fix $C_1$ and $C_2$.
We can write
\[\begin{array}{ll}
\one_{K_2}(|k+s|)T(k+s)-\one_{K_2}(|k|)T(k)=& (\one_{K_2}(|k+s|)-\one_{K_2}(|k|))T(k+s)\\[2mm]
-\one_{K_2}(|k|)(T(k+s)-T(k))=:T_1+T_2.
\end{array}\]
By dominated convergence $ \norme{T_1}^2_{{\frak{H}}} \fld 0 $ as
$s\fld 0$. Now
\[\begin{array}{ll}
\norme{T_2}^2_{{\frak{H}}}&=\int
\one_{K_2}(|k|)\norme{T(k+s)-T(k)}^2_{B(\cH)} \d k \leq
\int_{C_1/2}^{2C2}\norme{T(k+s)-T(k)}^2_{B(\cH)} \d k
\end{array}
\]
for $s<C_1/4$. Next we have
\[
\bear{l}
T(k+s)-T(k)=(\EN-\HNren-\omega(k))^{-1}(\vren(k+s)-\vren(k))\langle X\rangle^{-1}\\[2mm]
\quad+(\EN-\HNren-\omega(k))^{-1}(\EN-\HNren-\omega(k+s))^{-1}\vren(k+s)(\om(k+s)-\om(k))\langle
X\rangle^{-1}, \ear
\]
so
\[
\bear{l} \norme{T(k+s)-T(k)}_{B(\cH)}\leq
\frac{1}{\om(k)}\normebk{(\vren(k+s)-\vren(k))\langle
X\rangle^{-1}}\\[2mm]
\quad+ \frac{1}{\om(k)}\frac{1}{\om(k+s)} \normebk{(\vren(k+s)\langle X\rangle^{-1}}(|k+s|-|k|)\\[2mm]
\quad \leq C(C_1,C_2) \normebk{(\vren(k+s)-\vren(k))\langle
X\rangle^{-1}}+ C(C_1,C_2)|s|\normebk{\vren(k)\langle
X\rangle^{-1}}^2 \ear
\]
uniformly for $C_1/2<|k|<2C_2$ and $|s|<C_1/4$ where $C(C_1,C_2)$ is a constant which depends on $C_1$ and $C_2$.

\noindent Since, as one can easily verify, for arbitrary $0<D_1<D_2$
\[\lim_{s\fld 0}\int_{D_1<\modk<D_2}\norme{(\vren(k)-\vren(k+s))\langle X \rangle^{-1}}^2\d k=0,\]

\noindent we can conclude
\[\lim_{s\fld 0}\int_{C_1/2}^{2C2}\norme{T(k+s)-T(k)}^2_{B(\cH)} \d k=0.\]
By fixing first $C_1<<1$ and $C_2>>1$, letting then $s\fld 0$, the
proof is concluded.\qed\\

 We recall the following:
 \bep \label{recall pseudodiff} Let $f\in \l2(\rr^d, \d k; \mathcal{B})$ where $\mathcal{B}$ is a Banach space
 and let us denote $U_s$ the group of isometries given by $U_s f (k):= f(k+s)$. Suppose
$ \norme{f-U_s f}\fld 0 \quad as \quad s\fld 0$.
 Then, for any $F\in \coinf(\rr^d)$  such that $F(0)=1$, \[
\norme{1-F(\frac{D_k}{R})f}\fld 0 \quad as \quad R\fld \infty\]
where $F(\frac{D_k}{R})f=(2\pi)^{-d}\int \hat{F}(s)U_{-R^{-1}s}f
ds.$ \eep

\bel Let $F\in \coinf(\rr)$ be a cutoff function with $0\leq F\leq
1$, $F(s)=1$ for $|s|\leq \12$, $F(s)=0$ for $|s|\geq 1$. Let
$F_{R}(x)= F(\frac{|x|}{R})$. Then \[ \lim_{\mu\to 0,\, R\to
+\infty}(\psi_{\mu}, \d\G(1-F_{R})\psi_{\mu})=0. \]
\label{mainlemma} \eel
 \proof Set $\frak{H}:=L^{2}(\rrt, \d k; B(\cH)).$ As in \cite[Lemma 4.5]{g}, we obtain
\[
\begin{array}{l}
(\psi_{\mu}, \d\G(1-F_{R})\psi_{\mu}) \leq \|T(k)\|_\frak{H}
\|(1-F(\frac{|D_{k}|}{R})) T(k)\|_\frak{H}\normeH{\langle X
\rangle \psi_\mu}+ o(\mu^{0}).
\end{array}
\]

\noindent By Lemma \ref{lemma Tk}, $
\norme{T(k)-T(k+s)}_\frak{H}\fld 0$  as $s\fld 0$,
 hence by Proposition \ref{recall pseudodiff} \[\|(1-F(\frac{|D_{k}|}{R})) T(k)\|_\frak{H}\in
o(R^{0}).\]  So we can conclude that $ (\psi_{\mu},
\d\G(1-F_{R})\psi_{\mu})= o(R^0)+o (\mu^0).$ \qed\\

\noindent\textbf{Proof of Theorem \ref{main arai}:} as in
\cite[Theorem 1]{g}, by replacing the compact operator $\chi(N\leq
\lambda)\chi(H_{0}\leq \lambda)\G(F_{R})$ (where $\chi$ is a
smoothed characteristic function of the unit ball) by the compact
operator $\chi(N\leq \lambda)\chi(H_{0}\leq
\lambda)\G(F_{R})\chiRp(|X|)$, and using in addition that, as a
consequence of Corollary \ref{cor expdecay}, for any $\delta>0$ we
can choose $P$ large enough such that
$\|(1-\chi_P)(|X|)\psi_{\mu}\|\leq\delta,\label{proof7}$ uniformly
in $\mu$ for $\mu<\mu_0$. \qed

\appendix

\section{Appendix A}

In this section we give a sketch of the proof of Lemma
\ref{localiz cor}. The idea behind the proof is to compare
$\HNcmu$ with an auxiliary Hamiltonian where the electrons are
localized in some regions, and the photons are localized
near the electrons.\\

We recall the following fact about existence of some partitions of
 unity.

\bep \label{cluster}There exists a family of functions
$F_a:\rr^{3N}\fld\rr$, for ${a\subset\{1,...,N\}}$ such that
\renewcommand{\labelenumi}{(\roman{enumi})}
\begin{enumerate}\item $\sum_a F_a^2=1$,
 \item  \textit{for all $a\neq\emptyset$   }
  \textit{$\supp F_a \subset$\{ $X\in \rrtn \:|\: |X|\geq 1,  \min_{i\in a^c\, j\in a}(|\xf_i-\xf_j|,|\xf_j|)\geq
   C\}$ where $C$ is some positive constant,}
   \item \textit{if $a=\emptyset$, then $\supp \,F_a$ is
   compact},
\item \textit{let $\diesis a $ be the cardinality of the set $a$;
the functions $\sum_{\diesis a=p}F_a^2$ are symmetric for all
$0\leq p \leq N$}.
 \end{enumerate}
\eep

\proof see for example \cite{dege}.\qed\\

\noindent With this notation the subset $a$ will represent the
particles far from the origin. \\
 Each function of the family will be used to
localize fermions. Corresponding to each \fermion localization we
now define  \boson localization. For a given $a$, consider the
function
 \bdisp
\bear{ll}
 g_{\infty,\,a,\,P}(\xb,\X):=&\left\{
 \bear{lll}\prod_{j\in a}1-\chi(\frac{x-\xf_j}{P})&\quad\quad
 &a\neq\emptyset\\
 \chi(\frac{\xb}{P})&&a=\emptyset\\\ear \right.\\[2mm]
g_{0,\, a, \,P}(\xb,\X):=&1-g_{\infty,\,a}(\xb,\X)\\\ear
 \eedisp
 where
$\chi$ is a smoothed characteristic function of the unit ball.

\noindent Now let us set for $\eps=0,\infty :$ \bdisp
j_{\eps,\,a,\,P}:=j_{\eps,a}(\xb,\X):=
\frac{g_{\eps,\,a,\,P}(\xb,\X)}{\sqrt{g_{\infty,\,a,\,P}(\xb,\X)^2+g_{0,a,\,P}(\xb,\X)^2}}
\eedisp \noindent so that
$j_{0,\,a,\,P}^2+j_{\infty,\,a,\,P}^2=1$.
\begin{rmk}\label{supportij}
 Note that for $a\neq\emptyset$
 $$
 \bear{lcl}
 \supp j_{\infty, \, a,\,P} & \subset & \{ \xb\in\rrt \; | \;  |\xb-\xf_j|>
 P, \mbox{ for all }
j\in a\}\\
\supp j_{0, \, a,\, P}& \subset & \{ \xb\in\rrt \; | \;
|\xb-\xf_j|\leq P, \mbox{ for some } j\in a\}\\ \ear. $$
\end{rmk}

\vspace{3mm}

\noindent For each $a$ we define \beq \label{defKa} \bear{l}
K_a:=K-(\sum_{i\in a, j\notin a}w(\xf_i-\xf_j) + \sum_{i\in a}
v(\xf_j)).\ear
 \eeq
\noindent Now we define the cluster Hamiltonian $\Hcamu$ by
\beq\label{eq:defHam} \Hcamu:=K_a \ot \one+\one
\ot\dgommu+\lambda\Phi(\vmu). \eeq

The next lemma follow easily from hypothesis \textit{(I)}.
 \bel \label{localizzel} Let $\HNcmu$ be the Hamiltonian defined in (\ref{defH}).
 Assume (I). Then $F_{a,R} (\HNcmu-\Hcamu)=O(R^{-\eps})$ for all $a$, where $\eps:=inf\{\eps_1, \eps_2\}$.\eel

In order to deal with photon localization, we need to introduce
the \emph{extended cluster Hamiltonians} $\Hamext$. We introduce
the space $ \Hextsp:=\cK\otimes\gammash\otimes\gammash, $  on
which we define the following operators:
\[
\dgext:=\onek \otimes \dgommu\otimes\onegh +\onek \otimes \onegh
\otimes\dgommu,\]
\[
\phivaext:=\sumjna \Phi(\vmuj)\ot\onegh+\onegh\otc\sumja
\Phi(\vmuj),
\]
where $\vmuj:=\vmu(\xf_j, k)$.\\
\noindent We define
\[
\Hamext:=K_a\ot\one\ot\one+\one\ot\dgext+\lambda \phivaext
\]
The extended cluster Hamiltonians are built \emph{ad hoc} in order
to have, up to identifications,

\[\bear{lll}\Hamext & =&H_{N'\!,\;\mu}\ot\onea\ot\onegh+\oneac\ot\onegh\ot H_{N-N'\!,\;\mu}\\
\ear
\]

 \noindent where $\diesis a=N-N'$. This implies $
 \inf \si(\Hamext)=E_{N'\!,\;\mu}+E^0_{N-N'\!,\;\mu}$.\\

The following Lemma is well known.
 \bel \label{bound N by Hmu} Let
$\HNcmu$ be the Hamiltonian defined in (\ref{defHNcmu}). Then
there exist some constants $C,D \in \rr^+$ such that
\[
\cN\leq \frac{C \HNcmu +D}{\mu}.
\]
\eel

 \bel \label{localizzphot}
Let $\Hcamu$ the cluster Hamiltonian defined in (\ref{eq:defHam}).
Let $\jarp:=(\jinfarp,\joarp)$. Then
\renewcommand{\labelenumi}{(\roman{enumi})}
\begin{enumerate}
    \item if $a\neq\emptyset$,
    \[\Far(\Hcamu - \gca (\jarp) \Hamext
\gc(\jarp))=O(\frac{\ln^{\12}\!\mu}{\mu
    P})(\HNcmu+C)
    \]
    when $R=\gamma P$ with $\gamma>>1$,

    \item if $a=\emptyset$, the same holds   when $P=\gamma R$ with $\gamma>>1$.
\end{enumerate}
\eel

\proof \textit{(i)}  We have to evaluate
 \[\bear{l}
\Far\left(\Hcamu - \gca (\jarp) \Hamext \gc(\jarp)\right)
=\Far\left(K_a\ot\onegh+ \gca(\jarp)(K_a\ot\onegh\ot\onegh)
\gc(\jarp)\right)\\[2mm]

\quad+\Far \left(\dgommu-\gca(\jarp)\dgmext \gc(\jarp)\right) +
\lambda \Far \left(\phivm-\gca(\jarp)\phivamext
\gc(\jarp)\right)\\[2mm]\quad=:I_1+I_2+I_3.\ear\]

\noindent Note that $\jarp$ is a function of \emph{both}   $\X$ and $\xb$.\\[2mm]
Consider first $I_1$. Using (\ref{iso2}) we have:
\[
\begin{array}{ll}
  I_1&= \Far \left(\sumiuN \dg(ad^2_{\jorp} \frac{1}{2}\lapxi+ad^2_{\jinrp} \frac{1}{2}\lapxi)\right) \\[2mm]
&\leq \sumiuN \norme{ad^2_{\jorp} \frac{1}{2}\lapxi+ad^2_{\jinrp}
\frac{1}{2}\lapxi}_{\B(\cK\otimes\ch)}(\cN+1)
  \leq O({(\mu P)}^{-1})(\HNcmu+C)\\[2mm]
\end{array}
\]

\noindent by Lemma \ref{bound N by Hmu}.\\
\indent Consider now $I_2$. Using (\ref{iso2}), we obtain
$$
I_2\leq \norme{ad _{\jorp}^2 \ommu+
ad_{\jinrp}^2}_{\B(\ch)}(\cN+1)\leq O({(\mu P)}^{-1})(\HNcmu+C)$$

\noindent by Lemma \ref{bound N by Hmu}.\\
\indent Consider now $I_3$. Using (\ref{iso1}), it is easy to
compute
\[
\bear{ll} I_3&= \lambda \Far \left( \gca(\jarp)\left(\sumjna\Phi(
(\jinrp-1)\vmuj
)\otimes \onegh + \onegh \hat{\otimes} \Phi(\jorp  \vmuj)\right.\right.\\[2mm]&\left.\left.
+\sumja\Phi(\jinrp\vmuj)\otimes \onegh + \onegh \otimes
\Phi((\jorp-1) \vmuj)\right) \gc(\jarp)\right).\ear
\]
A term of the form $\tilde{\jmath}\;\vmuj$ (where $\tilde{\jmath}$
will be $\jorp$ , $\jinrp-1$, etc)  can be seen in two ways: as an
element of $\ch:=\l2(\rrt, \d k)$,  in this case $\tilde{\jmath}$
is a pseudodifferential operator on $\ch$, in other words
$\tilde{\jmath}\;\vmuj= \tilde{\jmath}\;(D_k)\vmu(\xf_j,k)$; or as
an element of $\ch_x:=\l2(\rrt, \d x)$, in this case we mean
$\tilde{\jmath}\;\vmuj=\tilde{\jmath}\;(x)v_\mu(x-\xf_j)$ with
$v_\mu(x-\xf_j):=\mathcal{F}\vmu(\xf_j,k)$  where $\mathcal{F}$ is
the Fourier transform with respect to the variable $k$. Anyway, by
unitary of $\mathcal{F}$,
$\normeh{\tilde{\jmath}\;\vmuj}=\norme{\tilde{\jmath}\;\vmuj}_{\ch_x}$
(see also the proof of Lemma \ref{Lemma pseudodiff2}), so we can
write $\norme{\tilde{\jmath}\;\vmuj}$ without ambiguity.

 \noindent Let's consider the terms of the form $A:=\Far \gca(\jarp)\ad\left( \tilde{\jmath}\;\vmuj\right)\ot
\one_{\gammash}\gc(\jarp)$ (here $\lambda$ is neglected). For
$u\in\H$
\[
\bear{l} \normeH{    \Far \gca(\jarp)\ad( \tilde{\jmath}\; \vmuj)\ot \onegh \gc(\jarp) u }^2\\[2mm]
\leq \int_{\supp \Far}  \norme{(\ad( \tilde{\jmath}\;\vmuj)\ot \onegh)(\cN^{{\rm ext}}+1)^{-1/2}\gc(\jarp) (\cN+1)^{1/2}u(X) }_{\gammash \ot \gammash}^2\d \X\\[3mm]
\leq \int_{\supp \Far}  \normeh{ \tilde{\jmath}\;\vmuj}^2 \langle u (\X),(\cN+1)u(\X)\rangle_{\gammash} \d \X \\
\ear
\]
since $ \norme{\ad(h)u}^2\leq \normeh{h}^2 (u, (\cN+ 1)\, u\,)$.\\[2mm]
The same estimate holds for the terms of the form
$
A':=\Far \gca(\jarp)\one_{\gammash}\ot\ad\left(
\tilde{\jmath}\;v_j\right)\gc(\jarp).
$

\noindent Hence we have to estimate the norms:\\

\begin{tabular}{lll}
   $\norme{\jinrp\vmuj},$ & $\norme{(\jorp-1)\vmuj}$ & \small{for all
$j\in a$, for $X\in \supp
\Far$}\\[2mm]
$\norme{(\jinrp-1)\vmuj},$&$\norme{\jorp\vmuj}$ & \small{for all
$j\notin a$, for $X\in \supp \Far$}\\[2mm]
\end{tabular}

 \noindent Since
$\supp j_{\infty, \, a,\,P} \subset \{ \xb\in\rrt \; | \;
|\xb-\xf_j|>
 P, \mbox{ for all }
j\in a\}$, then
 by Lemma \ref{Lemma pseudodiff2}$\norme{ \jinrp\vmuj}=O(\ln ^{\! \12}\!\!\mu\; P^{-1})$
 uniformly in $\xf_j$. The same holds for $\norme{(\jorp-1)\vmuj}$.

\noindent Now $\supp j_{0,\,a,\,P} \subset  \{ \xb\in\rrt \; | \;
|\xb-\xf_i|\leq P, \mbox{ for some } i\in a\}$
 but on the other hand $\X\in\supp\Far$ implies
 $|\xf_j-\xf_i|>R$. Choosing $R=\gamma
 P$ with $\gamma>>1$, we obtain, for $j\notin a$ and $\X \in\supp\Far $,
  $|\xb-\xf_j|>(\gamma -1)P$,  $\supp j_{0,\,a,\,P} \subset  \{ \xb\in\rrt \; | \; |\xb-\xf_j|\leq
(\gamma-1)P \}$.
  Hence by Lemma \ref{Lemma pseudodiff2}
$\norme{ \jorp\vmuj}^2=O(\ln\mu^{\12} \; P^{-1})$ uniformly in
$\xf_j$; the same holds for $\norme{(\jinrp-1)\vmuj}$.
 Collecting the estimates for $I_1$, $I_2$, $I_3$ we obtain the
 lemma.\\

\textit{(ii)} We proceed in the same way. Since $a=\emptyset$, we
only have to evaluate  norms of the type $\norme{\jorp\vmuj}$ . In
this case $\jinrp$ is compactly supported and $\supp j_{0, \,
a,\,P} \subset  \{ \xb\in\rrt \; | \; |\xb|> P\}$, but also $\Far$
is compactly supported \ie $|\X|< R$. Hence, in this case we have
to choose $R<<P$, for example $P=\gamma
 R$ with $\gamma>>1$ so that $|\xb-\xf_j|>\frac{(\gamma
 -1)}{\gamma}P$, hence by Lemma \ref{Lemma pseudodiff2} $\norme{ \jorp\vmuj}=O(\ln ^{\12}\mu \;
 P^{-1}).$\qed\\[2mm]

\noindent \textbf{Corollary [Lemma \ref{localiz cor}]}\textit{ Let
$\HNcmu$ be the Hamiltonian defined in (\ref{defH}). Then \[
\HNcmu\geq\tau_\mu-f(\mu)o\left(\, R^{0}\right)(\HNcmu+C) \hbox{
on  } \DR,
\]
where $f(\mu):=\frac{\ln^{\12} \mu}{\mu}$ and  $\DR:= \{\psi \in
\cH\,|\, \psi(X)=0 \mbox{ if } |X|<R\}.$}\\
 \noindent \proof Easy
using the previous Lemma and IMS localization formula. See
\cite{gll}.\qed

\section{Appendix B}
 \bet[functional calculus formula]\label{almost analytic} Let
$f\in\coinf(\rr)$ and $H$ be a self-adjoint
 operator  on a Hilbert space, then there  exists
a function $\ft\in\coinf(\C)$ such that $\ft_{|\rr}=f$,
$\module{\dd{\ft}{\bar{z}}}\leq c_n \module {\Im
 z}^n$ for all $n\in \N$ and  \bdisp
f(H)=\frac{1}{\ic\pi}\int_{\C} \dd{\ft}{\bar{z}}(z)(z-H)^{-1}\dz.
\eedisp \eet

 The function $\ft$ is called an \emph{almost-analytic extension}
of
$f$ .\\
\proof See for instance \cite{helf}.\qed

\end{document}